\documentclass[letterpaper,twocolumn,english,prl,superscriptaddress]{revtex4-1}
\usepackage{amsmath}
\usepackage{newtxmath}
\usepackage[T1]{fontenc}
\usepackage[latin9]{inputenc}
\usepackage{color}
\usepackage{babel}
\usepackage{calc}
\usepackage{mathrsfs}
\usepackage{mathdots}
\usepackage{stackrel}
\usepackage{graphicx}
\usepackage[unicode=true,pdfusetitle,
 bookmarks=true,bookmarksnumbered=false,bookmarksopen=false,
 breaklinks=false,pdfborder={0 0 1},backref=false,colorlinks=true]
 {hyperref}



\begin{document}
\title{Measurement-induced integer families of critical dynamical scaling
in quantum many-body systems}
\author{Zuo Wang}
\affiliation{Key Laboratory of Atomic and Subatomic Structure and Quantum Control
(Ministry of Education), School of Physics, South China Normal University,
Guangzhou 510006, China}
\affiliation{Guangdong Provincial Key Laboratory of Quantum Engineering and Quantum
Materials, Guangdong-Hong Kong Joint Laboratory of Quantum Matter,
South China Normal University, Guangzhou 510006, China}
\author{Shi-Liang Zhu}
\email{slzhu@scnu.edu.cn}

\affiliation{Key Laboratory of Atomic and Subatomic Structure and Quantum Control
(Ministry of Education), School of Physics, South China Normal University,
Guangzhou 510006, China}
\affiliation{Guangdong Provincial Key Laboratory of Quantum Engineering and Quantum
Materials, Guangdong-Hong Kong Joint Laboratory of Quantum Matter,
South China Normal University, Guangzhou 510006, China}
\author{Li-Jun Lang}
\email{ljlang@scnu.edu.cn}

\affiliation{Key Laboratory of Atomic and Subatomic Structure and Quantum Control
(Ministry of Education), School of Physics, South China Normal University,
Guangzhou 510006, China}
\affiliation{Guangdong Provincial Key Laboratory of Quantum Engineering and Quantum
Materials, Guangdong-Hong Kong Joint Laboratory of Quantum Matter,
South China Normal University, Guangzhou 510006, China}
\author{Liang He}
\email{liang.he@scnu.edu.cn}

\affiliation{Key Laboratory of Atomic and Subatomic Structure and Quantum Control
(Ministry of Education), School of Physics, South China Normal University,
Guangzhou 510006, China}
\affiliation{Guangdong Provincial Key Laboratory of Quantum Engineering and Quantum
Materials, Guangdong-Hong Kong Joint Laboratory of Quantum Matter,
South China Normal University, Guangzhou 510006, China}
\begin{abstract}
A quantum many-body system can undergo transitions in the presence
of continuous measurement. In this work, we find that a generic class
of critical dynamical scaling behavior can emerge at these measurement-induced
transitions. Remarkably, depending on the symmetry that can be respected
by the system, different integer families of dynamical scaling can
emerge. The origin of these scaling families can be traced back to
the presence of hierarchies of high order exceptional points in the
effective non-Hermitian descriptions of the systems. Direct experimental
observation of this class of dynamical scaling behavior can be readily
achieved using ultracold atoms in optical lattices or through intermediate-scale
quantum computing systems. 
\end{abstract}
\maketitle
\emph{Introduction}.\textemdash Quantum technology has undergone rapid
development in recent decades \citep{Preskill_Quantum_2018}. Particularly,
in the field of quantum information science, both digital quantum
computers, represented by quantum circuits \citep{lloyd1996science,arute2019nature,zhong2020science,wu2021PRL},
and analog quantum computers, represented by programmable quantum
simulators \citep{georgescu2014RMP,bernien2017nature,browaeys2020nature,semeghini2021nature,giudici2022PRL},
have achieved high levels of controllability. Quantum measurements
play a crucial role in extracting useful information from these devices,
and in certain cases, they can also be utilized to prepare nontrivial
quantum states, such as random states that are essential for quantum
information applications \citep{Choi_Nat_Phys_2023}. Recently, these
technological advancements have also motivated investigations into
a fundamental class of questions, broadly categorized as measurement-induced
transitions in quantum many-body systems. These investigations encompass
transitions induced by different types of measurement protocols, including
continuous measurements \citep{Gullans2020PRX,Shao-Kai2021PRL,Fuji2020PRB},
random projective selections \citep{Skinner2019PRX,Yaodong2018PRB,Yaodong2019PRB,Fuji2020PRB,Ruihua2021PRB},
measurements with post-selection \citep{Nahum2021PRX,gopalakrishnan_Gullans2021PRL,Jian2021PRB,Turkeshi2021PRB}.
Notably, non-trivial dynamical scaling behavior of entanglement has
been identified in various scenarios \citep{Chan2019PRB,Yaodong2019PRB,Choi2020PRL,Gullans2020PRX,Goto2020PRA,Chao-Ming2020PRB,Fuji2020PRB,Gullans2020PRL,Ruihua2021PRB,Shao-Kai2021PRL,Sang2021PRR,noel2022nature,Skinner2019PRX,Yaodong2018PRB,Nahum2021PRX,gopalakrishnan_Gullans2021PRL,Jian2021PRB,Turkeshi2021PRB}.

Viewing measurements as influences imposed by the environment, quantum
systems under measurement fall into the category of open quantum systems
\citep{Gardiner_2000_Springer,Breuer_2000_Oxford}. These systems
exhibit key features such as possible breaking of detailed balance,
making them inherently non-equilibrium. From this perspective, measurement-induced
transitions can be classified as non-equilibrium transitions, which
can exhibit much richer dynamical scaling behavior compared to equilibrium
transitions. For instance, the well-known dynamical behavior described
by the Kardar-Parisi-Zhang equations is not only found in typical
growing interface dynamics \citep{kardar1986PRL} but also observed
in various quantum systems such as exciton-polariton condensates \citep{deligiannis2022PRR,fontaine2022nature}.
Remarkably, certain families of dynamical scaling can emerge in non-equilibrium
systems, such as the Fibonacci family of dynamical scaling found in
the multi-component asymmetric simple exclusion process \citep{popkov2015PNAS}.
In this context, a fundamental question arises regarding the existence
of non-trivial critical dynamical scaling in physical observables
and their properties beyond the entanglement of the systems.

In this work, we address this question for several quantum many-body
systems under continuous measurements and find that a generic class
of dynamical scaling indeed emerges at these measurement-induced transitions.
Remarkably, different integer families of dynamical scaling (IFDS)
can emerge depending on the symmetry that can be respected by the
system. These scalings are distinct from the scaling of conventional
phase transitions as they can emerge even in finite-sized systems
without requiring the thermodynamic limit. The origin of these scaling
families can be attributed to the hierarchies of exceptional points
(EPs) present in the effective non-Hermitian descriptions of the systems.
Experimental observation of this class of dynamical scaling can be
readily performed in various current experimental setups.

\begin{figure}
\includegraphics[width=3.3in]{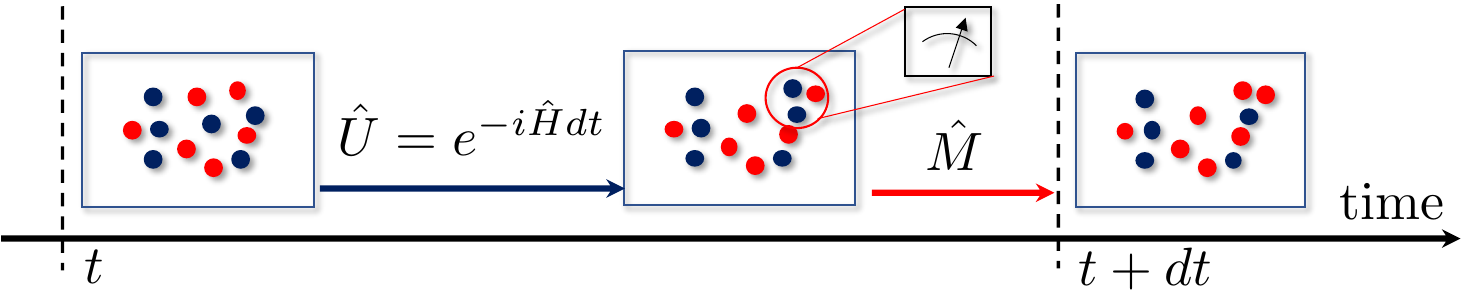}

\caption{\label{fig:Dynamics_illustration}Schematic illustration of the quantum
many-body dynamics in the presence of continuous measurements. In
each short time interval $dt$, the system undergoes a unitary evolution
$\hat{U}=e^{-i\hat{H}dt}$ governed by its Hamiltonian $\hat{H}$.
Subsequently, a local or global measurement $\hat{M}$ is conducted
on the system, and the quantum trajectories are post-selected based
on the measurement outcome.}
\end{figure}

\emph{Measurement protocol and measurement-induced dynamical scaling
in the transverse Ising model}.\textemdash The type of the measurement
protocol considered here pertains to a continuous monitoring or variable
strength weak measurements \citep{Ivanov_PRL_2020}. For a generic
quantum many-body system, its dynamics within each short time interval
$dt$ is governed by its Hamiltonian $\hat{H}$ and an operator $\hat{M}$
that encapsulates the measurement effect on the system. The evolution
of the quantum state $|\psi(t)\rangle$ is given by $|\psi(t+dt)\rangle=\hat{M}e^{-i\hat{H}dt}|\psi(t)\rangle$
(units $\hbar=1$) as illustrated schematically in Fig.~\ref{fig:Dynamics_illustration}.

Let us begin by considering the transverse Ising model under monitoring
(the periodic boundary is assumed). Previous studies have demonstrated
that this model, under global monitoring, exhibits a measurement-induced
transition \citep{Biella2021Quantum}. Here, we explore a more general
scenario where the transverse field is local, and the measurement
is also performed on the region of the local transverse field. The
system's dynamics is determined by the Hamiltonian $\hat{H}_{\mathrm{TI}}$
and the measurement operator $\hat{M}_{\text{S}}$, where 
\begin{align}
\hat{H}_{\mathrm{TI}} & =-J\sum_{j=1}^{L}\hat{\sigma}_{j}^{z}\hat{\sigma}_{j+1}^{z}-g\sum_{j\in\mathcal{M}}\hat{\sigma}_{j}^{x},\label{eq:Transverse_Ising}\\
\hat{M}_{\text{S}} & =1-\gamma dt\sum_{j\in\mathcal{M}}(1-\hat{\sigma}_{j}^{y}).\label{eq:Spin_Measurement}
\end{align}
Here, $\hat{\sigma}_{j}^{x}$, $\hat{\sigma}_{j}^{y}$, $\hat{\sigma}_{j}^{z}$
represent the three Pauli operators on site $j$, $J$ denotes the
strength of the ferromagnetic coupling, $g$ represents the strength
of the local transverse field along the $x$-direction, and $\mathcal{M}$
denotes both region of measurement and local transverse field. The
measurement effect on the system is encapsulated in the operator $\hat{M}_{\mathrm{S}}$.
Specifically, it corresponds to performing a weak measurement on the
spin-down state along the $y$-direction with a strength characterized
by $\gamma$, and post-selecting the trajectories where the spin-down
state along the $y$-direction is not detected \citep{Biella2021Quantum}
(see Supplemental Material (SM) \citep{Sup_Mat}).

Prior investigations have shown that when the measurement strength
$\gamma$ matches the strength of the transverse field $g$, the entanglement
of the system undergoes a sharp transition and exhibits critical behavior
\citep{Biella2021Quantum}. Now, we examine whether such criticality
is also manifested in other experimentally accessible observables.

\begin{figure}
\includegraphics[width=1.6in]{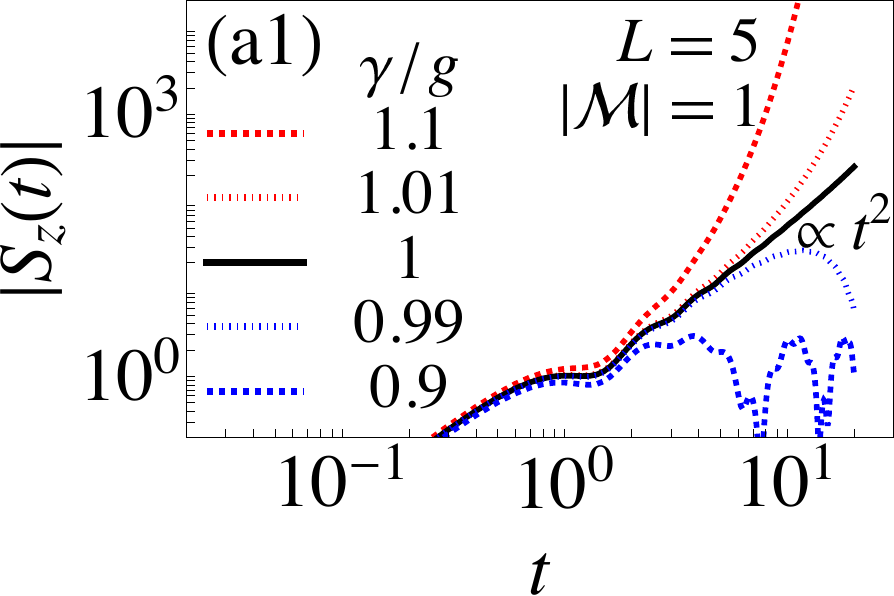}~\includegraphics[width=1.6in]{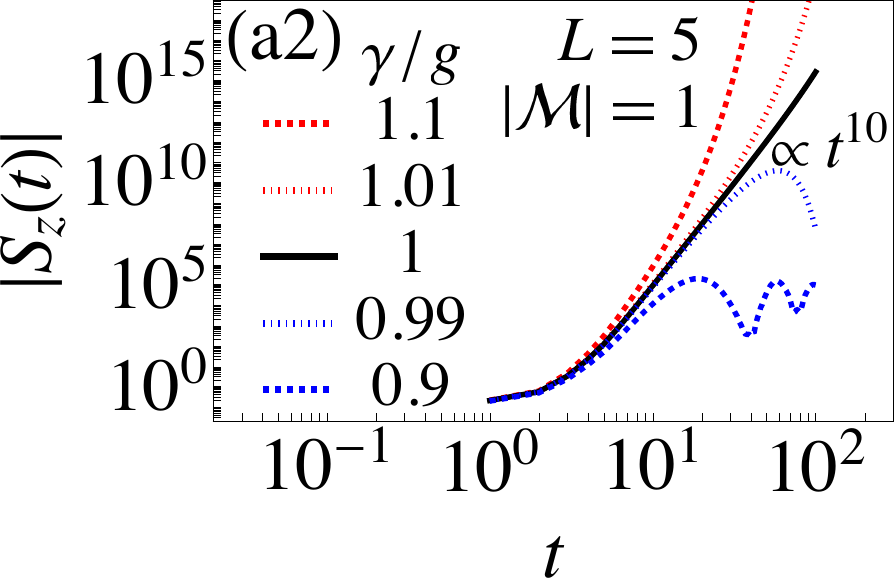}

\includegraphics[width=1.6in]{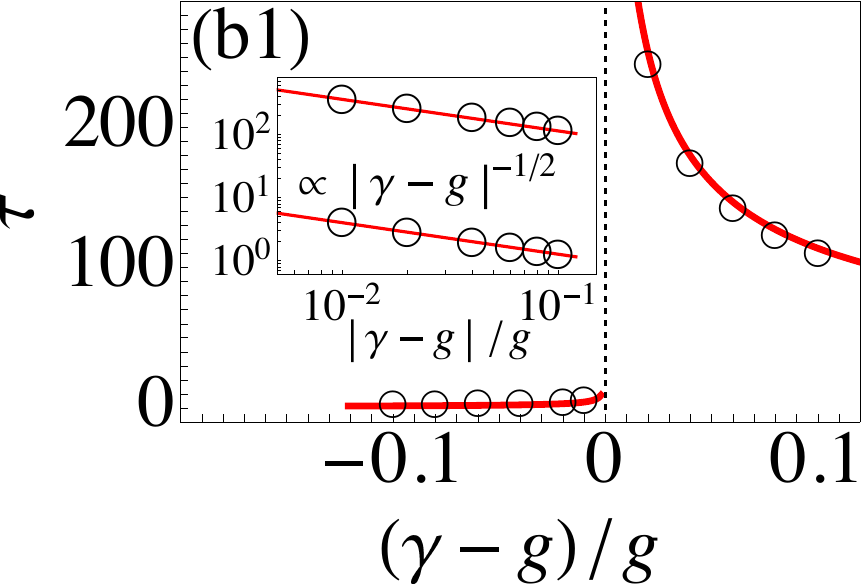}~\includegraphics[width=1.6in]{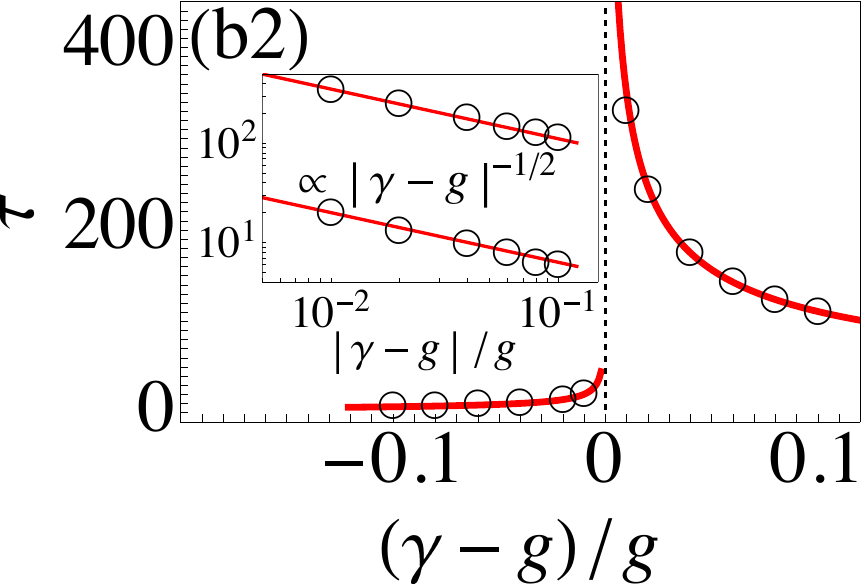}

\includegraphics[width=1.6in]{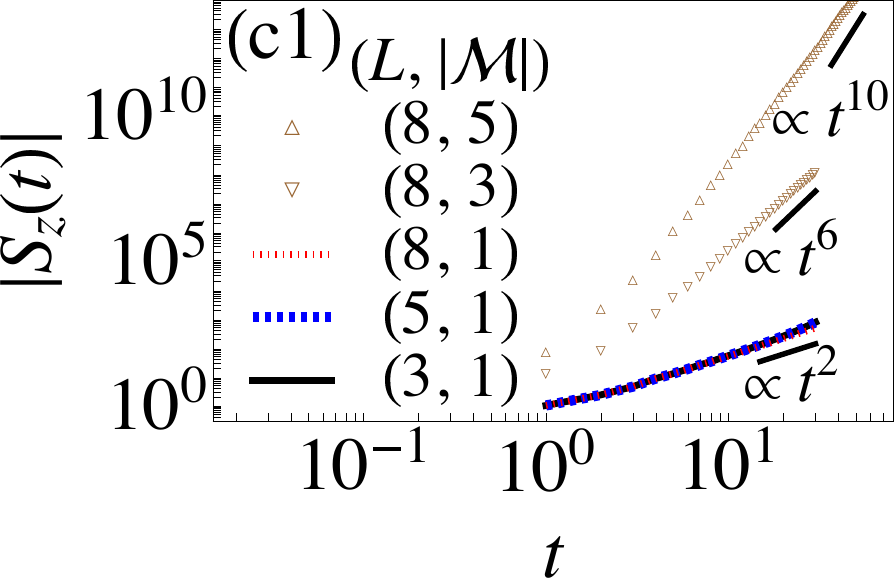}~\includegraphics[width=1.6in]{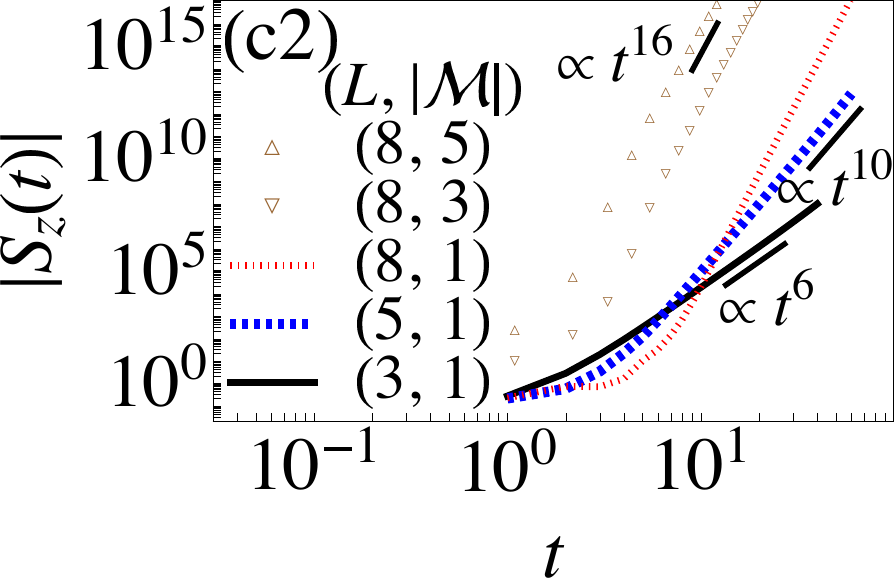}

\caption{\label{fig:Scaling_Spin_Models}Quantum dynamics of the transverse
Ising (a1,~b1,~c1) and Heisenberg (a2,~b2,~c2) model under monitoring.
(a1,~a2) Time dependence of $|S_{z}(t)|$ at different $\gamma/g$
values. Power law scaling $|S_{z}(t)|\propto t^{2}$ emerges as $\gamma/g\rightarrow1$.
(b1,~b2) Characteristic time $\tau$ manifests a power divergence
$\tau\propto|\gamma-g|^{-1/2}$ for $\gamma/g\rightarrow1$. (c1,~c2)
Power law scaling of $|S_{z}(t)|$ at critical point $\gamma/g=1$
for different system and measurement region sizes. Transverse Ising
model manifests an IFDS $|S_{z}(t)|\propto t^{2|\mathcal{M}|}$ (c1),
while transverse Heisenberg model manifests an IFDS $|S_{z}(t)|\propto t^{2L}$
(c2). All dynamics are simulated with $J=1$, $g=1$, $dt=10^{-4}$,
$\mathcal{M}=\{1,2,\cdots,|\mathcal{M}|\}$, and the initial state
give by $|\psi(t=0)\rangle=2^{-L}\sum_{\{\sigma_{i}\}}|\sigma_{1}\sigma_{2}\cdots\sigma_{L}\rangle$.}
\end{figure}

To this end, we numerically simulate the dynamics according to $|\psi(t+dt)\rangle=\hat{M}_{\text{S}}e^{-i\hat{H}_{\mathrm{TI}}dt}|\psi(t)\rangle$
and calculate the time dependence of the total magnetization along
the $z$-direction $|S_{z}(t)|\equiv|\langle\Psi(t)|\sum_{j=1}^{L}\hat{\sigma}_{j}^{z}/2|\Psi(t)\rangle|$
in the ``renormalized'' state $|\Psi(t)\rangle\equiv e^{|\mathcal{M}|\gamma t}|\psi(t)\rangle$
with $|\mathcal{M}|$ being the cardinality of $\text{\ensuremath{\mathcal{M}}}$.
The pre-factor $e^{|\mathcal{M}|\gamma t}$ introduced in $|\Psi(t)\rangle$
is intended to eliminate the trivial exponential decay of the module
of $|\psi(t)\rangle$ caused by the post-selection of trajectories.
Fig.~\ref{fig:Scaling_Spin_Models}(a1) illustrates the time evolution
of $|S_{z}(t)|$ for a system with $L=5$ and $|\mathcal{M}|=1$ at
different measurement strengths $\gamma$. Notably, as $\gamma$ approaches
$g$, $|S_{z}(t)|$ manifests a scaling behavior of $|S_{z}(t)|\propto t^{2}$
in the ``late time'' dynamics.

To quantitatively analyze the emergence of this scaling, we determine
a characteristic timescale $\tau$ by performing fits of a simple
exponential function $Ae^{t/\tau}$ and an oscillatory function $B\cos(t/\tau+\theta)+C$
($A$, $B$, $C$, $\theta$ and $\tau$ are real fitting parameters)
to the numerical data of $|S_{z}(t)|$ for $\gamma/g>1$ and $\gamma/g<1$,
respectively. As shown in Fig.~\ref{fig:Scaling_Spin_Models}(b1),
the emergence of the dynamical scaling coincides with the divergence
of the characteristic timescale $\tau$, reminiscent of the similar
behavior in traditional continuous phase transitions. Interestingly,
we observe a power law scaling of $\tau$ with respect to the distance
between the measurement strength $\gamma$ to its critical value $\gamma=g$,
i.e., $\tau\propto|\gamma-g|^{-1/2}$. Furthermore, for this single-site
measurement case ($|\mathcal{M}|=1$), we calculate $|S_{z}(t)|$
at the transition point $g=\gamma$ at different system sizes with
$L=3,5,8$. As shown by the lower three curves in Fig.~\ref{fig:Scaling_Spin_Models}(c1),
we observe that $|S_{z}(t)|$ manifests the same dynamical scaling
$\propto t^{2}$ in the ``late time'' dynamics. These results clearly
demonstrate that, beyond the entanglement entropy, other physical
observables can indeed exhibit critical scaling at the transition
point.

\emph{IFDS in transverse Heisenberg and Ising spin chain under monitoring}.\textemdash The
scaling behavior at the critical point of a system is known to be
strongly influenced by its underlying symmetry. Hence, it is of great
interest to explore systems that exhibit different symmetries. Motivated
by this, we investigate the behavior of a local transverse field ferromagnetic
spin-1/2 Heisenberg chain under monitoring, considering a periodic
boundary condition. The dynamics of this system are governed by the
Hamiltonian 
\begin{align}
\hat{H}_{\mathrm{TH}} & =-J\sum_{j=1}^{L}\sum_{a=x,y,z}\hat{\sigma}_{j}^{a}\hat{\sigma}_{j+1}^{a}-g\sum_{j\in\mathcal{M}}\hat{\sigma}_{j}^{x},\label{eq:Transverse_Heisenberg}
\end{align}
along with the measurement operator $\hat{M}_{\text{S}}$ defined
in Eq.~(\ref{eq:Spin_Measurement}). Fig.~\ref{fig:Scaling_Spin_Models}(a2)
illustrates the time evolution of $|S_{z}(t)|$ for the system with
$L=5$, $|\mathcal{M}|=1$ at various measurement strength $\gamma$.
Similar to the transverse Ising model, as $\gamma$ approaches $g$,
$|S_{z}(t)|$ manifests scaling behavior characterized by $|S_{z}(t)|\propto t^{\alpha}$
in its late-time dynamics. However, in contrast to the Ising case,
the scaling exponent assumes a surprisingly larger value of $\alpha=10$,
even though the characteristic timescale $\tau$ exhibits the same
diverging behavior, i.e., $\tau\propto|\gamma-g|^{-1/2}$ {[}Fig.~\ref{fig:Scaling_Spin_Models}(b2){]}.

To validate the value of the scaling exponent, we further examine
time dependence of $|S_{z}(t)|$ at the critical point for different
system sizes. However, we observe a strong dependence of the dynamical
scaling of $|S_{z}(t)|$ on the system size $L$, as shown in Fig.~\ref{fig:Scaling_Spin_Models}(c2).
At first glance, this may seem to be a manifestation of the finite
size effects. However, upon closer inspection of the scaling exponent
$\alpha$ for different system sizes, we find that they all take sharp
integer values. In fact, from the results presented in Fig.~\ref{fig:Scaling_Spin_Models}(c2),
one can deduce a simple empirical formula $\alpha=2L$ for the exponents
corresponding to different system sizes. Calculations involving more
system sizes (see SM \citep{Sup_Mat}) further confirm the validity
of this empirical formula.

Interestingly, for the transverse Ising chain under monitoring {[}see
Eqs.~(\ref{eq:Transverse_Ising},~\ref{eq:Spin_Measurement}){]},
although the scaling exponent does not change with respect to the
system size $L$, it actually manifest a dependence on the size of
the measurement region $|\mathcal{M}|$. As shown by the upper curves
in Fig.~\ref{fig:Scaling_Spin_Models}(c1) which correspond to different
$|\mathcal{M}|$, we observe that the exponent can be empirically
described by a simple formula $\alpha=2|\mathcal{M}|$. Further calculations
involving additional measurement sizes (see SM \citep{Sup_Mat}) support
the validity of this empirical formula.

The numerical results presented above strongly indicate that these
two systems can exhibit distinct IFDS, with their exponents being
determined either by the system size in the Heisenberg case or by
the size of the measurement region in the Ising case. Moreover, these
findings suggest that the emergence of these IFDS cannot be merely
coincidental. Indeed, as we shall show in the following, the emergence
of these IFDS can be attributed to the presence of series of high-order
EPs in the non-Hermitian effective descriptions of these systems.

\emph{Non-Hermitian effective descriptions}.\textemdash In the limit
of continuous measurement ($\gamma dt\ll1$), it is possible to construct
an effective non-Hermitian Hamiltonian \citep{gopalakrishnan_Gullans2021PRL,Turkeshi2021PRB,Jian2021PRB}
that faithfully captures (up to Trotter error) the system's physics
\citep{Sup_Mat}. For the transverse Ising chain under monitoring
{[}Eqs.~(\ref{eq:Transverse_Ising},~\ref{eq:Spin_Measurement}){]},
the measurement $\hat{M}_{\text{S}}$ can be well approximated by
$e^{-\gamma dt\sum_{j\in\mathcal{M}}(1-\hat{\sigma}_{j}^{y})}$ in
this limit. This allows us to effectively capture the system's time
evolution through the introduction of a non-Hermitian effective Hamiltonian,
denoted as $\hat{H}_{\mathrm{TI}}^{\mathrm{eff}}=-J\sum_{j=1}^{L}\hat{\sigma}_{j}^{z}\hat{\sigma}_{j+1}^{z}-\sum_{j\in\mathcal{M}}[g\hat{\sigma}_{j}^{x}+i\gamma(1-\hat{\sigma}_{j}^{y})]$.
Notably, the non-Hermitian term, $-i\sum_{j\in\mathcal{M}}\gamma(1-\hat{\sigma}_{j}^{y})$
adequately incorporates the physical consequences of the measurements.
Similar construction can be carried out for the Heisenberg spin-chain
{[}Eq.~(\ref{eq:Transverse_Heisenberg}){]} under monitoring, leading
to a non-Hermitian effective Hamiltonian $\hat{H}_{\mathrm{TH}}^{\mathrm{eff}}=-J\sum_{j=1}^{L}\sum_{a=x,y,z}\hat{\sigma}_{j}^{a}\hat{\mathbf{\sigma}}_{j+1}^{a}-\sum_{j\in\mathcal{M}}[g\hat{\sigma}_{j}^{x}+i\gamma(1-\hat{\sigma}_{j}^{y})]$.\textcolor{red}{}

We notice from Fig.~\ref{fig:Scaling_Spin_Models} that both for
the Ising and the Heisenberg cases, the onset of dynamical scaling
behavior is observed precisely at the parameter point $g=\gamma$.
Remarkably, this parameter point corresponds to the one at which the
EPs emerge in the local competition term between the transverse field
and the measurement, i.e., $g\hat{\sigma}_{j}^{x}+i\gamma(1-\hat{\sigma}_{j}^{y})$
in the non-Hermitian effective Hamiltonians $\hat{H}_{\mathrm{TI}}^{\mathrm{eff}}$
and $\hat{H}_{\mathrm{TH}}^{\mathrm{eff}}$. While it is generally
not guaranteed for a non-Hermitian Hamiltonian that the EP of its
constituent parts aligns with its own EP, we find that this is indeed
the case in this scenario (see SM \citep{Sup_Mat} for mathematical
proof). This intriguing outcome can be attributed to the symmetry
associated with the domain-wall conservation preserved by $\hat{H}_{\mathrm{TI}}^{\mathrm{eff}}$
and the $\text{SU}(2)$ symmetry respected by $\hat{H}_{\mathrm{TH}}^{\mathrm{eff}}$
at $g=\gamma=0$ (see SM \citep{Sup_Mat}). Specifically, in the Ising
case with $|\mathcal{M}|\leq L-2$, we find that EPs of different
orders appear for $\hat{H}_{\mathrm{TI}}^{\mathrm{eff}}$ at $g=\gamma$,
with the highest order being $|\mathcal{M}|+1$. While in the Heisenberg
case, regardless of the size of the measurement region $|\mathcal{M}|$,
EPs of different orders appear for $\hat{H}_{\mathrm{TH}}^{\mathrm{eff}}$
at $g=\gamma$, with the highest order being $L+1$. As we shall see
in the following, it is precisely these EPs of the highest order that
gives rise to the observed IFDS in the transverse Heisenberg and Ising
chain, as shown in Figs.~\ref{fig:Scaling_Spin_Models}(c1,~c2).

According to the general theory regarding the Jordan form \citep{bronson1991matrix},
for a generic non-Hermitian Hamiltonian $\hat{\mathcal{H}}$, some
of the eigenvectors belonging to the same eigenvalue coalesce at EPs.
Consequentially, in order to find a complete basis spanning the entire
Hilbert space, one can use the set of so-called generalized eigenvectors
$\{|V_{n}^{(j)}\rangle\}$ \citep{bronson1991matrix}, where $|V_{n}^{(j)}\rangle$
denotes the $n$th generalized eigenvector that corresponds to the
$j$th Jordan block with the eigenvalue $\lambda_{j}$. With this
complete basis $\{|V_{n}^{(j)}\rangle\}$, any generic state $|\psi\rangle$
can be expanded as $|\psi\rangle=\sum_{j=1}^{N_{\mathrm{JB}}}\sum_{n=1}^{\mathscr{O}_{j}}c_{n}^{(j)}|V_{n}^{(j)}\rangle$,
where $c_{n}^{(j)}$ are the expansion coefficients, $N_{\mathrm{JB}}$
is the number of Jordan blocks of $\hat{\mathcal{H}}$, $\mathscr{O}_{j}$
is the order of the $j$th Jordan block (for $\mathscr{O}_{j}\geq2$,
$\mathscr{O}_{j}$ is the order of the corresponding EP). Notably,
the time evolution of generalized eigenvectors assumes the form \citep{Ashida2020AP,Sup_Mat}
\begin{equation}
e^{-i\hat{\mathcal{H}}t}|V_{n}^{(j)}\rangle=e^{-i\lambda_{j}t}\sum_{m=0}^{n-1}\frac{(-it)^{m}}{m!}|V_{n-m}^{(j)}\rangle.\label{eq:GE_evo}
\end{equation}
From Eq.~(\ref{eq:GE_evo}), it becomes apparent that in the late-time
regime ($t\gg1$), the dominant term in $|\psi(t)\rangle$ is $e^{-i\lambda_{j_{\mathrm{max}}}t}(-it)^{\mathscr{O}_{j_{\mathrm{max}}}-1}[(\mathscr{O}_{j_{\mathrm{max}}}-1)!]^{-1}|V_{1}^{(j_{\mathrm{max}})}\rangle$,
where $\mathscr{O}_{j_{\mathrm{max}}}=\max(\{\mathscr{O}_{j}\})$.
This indicates that the expectation value of generic observables $\langle\hat{O}\rangle(t)\equiv\langle\psi(t)|\hat{O}|\psi(t)\rangle$
should exhibit the dynamical scaling $\propto t^{2(\mathscr{O}_{j_{\mathrm{max}}}-1)}$
in the late-time dynamics provided $\langle V_{1}^{(j_{\mathrm{max}})}|\hat{O}|V_{1}^{(j_{\mathrm{max}})}\rangle\neq0$
\citep{Remarks_More_IF}. As we have seen earlier, for $\hat{H}_{\mathrm{TH}}^{\mathrm{eff}}$
at $g=\gamma$, the EP with the highest order is of order $L+1$,
i.e., $\mathscr{O}_{j_{\mathrm{max}}}=L+1$, and the EP with the highest
order is of order $|\mathcal{M}|+1$ for $\hat{H}_{\mathrm{TI}}^{\mathrm{eff}}$
at $g=\gamma$. Consequently, in the late-time regime of the Heisenberg
and Ising case, the expectation value of generic observables, such
as $|S_{z}(t)|$, is expected to exhibit dynamical scaling behavior
of $\propto t^{2L}$ and $\propto t^{2|\mathcal{M}|}$, respectively.
This precisely corresponds to the observed IFDS in the transverse
Heisenberg and Ising chain under monitoring {[}see Figs.~\ref{fig:Scaling_Spin_Models}(c1,~c2){]}.
Furthermore, the power law divergence of the characteristic timescales
$\tau$ at the measurement induced transition {[}see Figs.~\ref{fig:Scaling_Spin_Models}(b1,~b2){]}
can also be attributed to the dynamical behavior in the vicinity of
the EPs associated with the parity-time reversal symmetry breaking
of the systems' non-Hermitian effective descriptions (see SM \citep{Sup_Mat}).

\begin{figure}
\includegraphics[width=1.6in,height=1.08in]{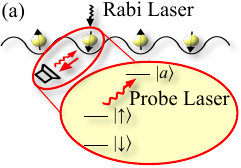}~\includegraphics[width=1.6in]{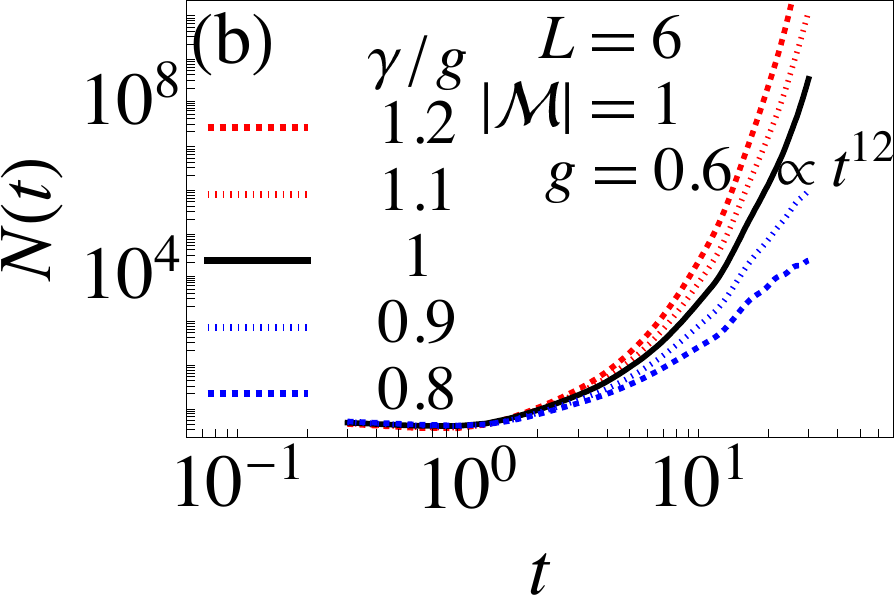}

\includegraphics[width=1.6in]{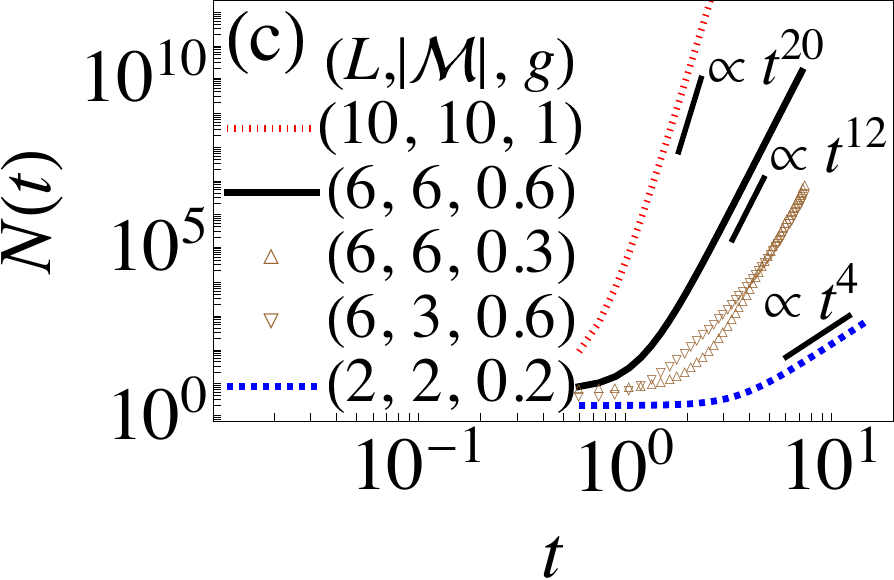}~\includegraphics[width=1.6in]{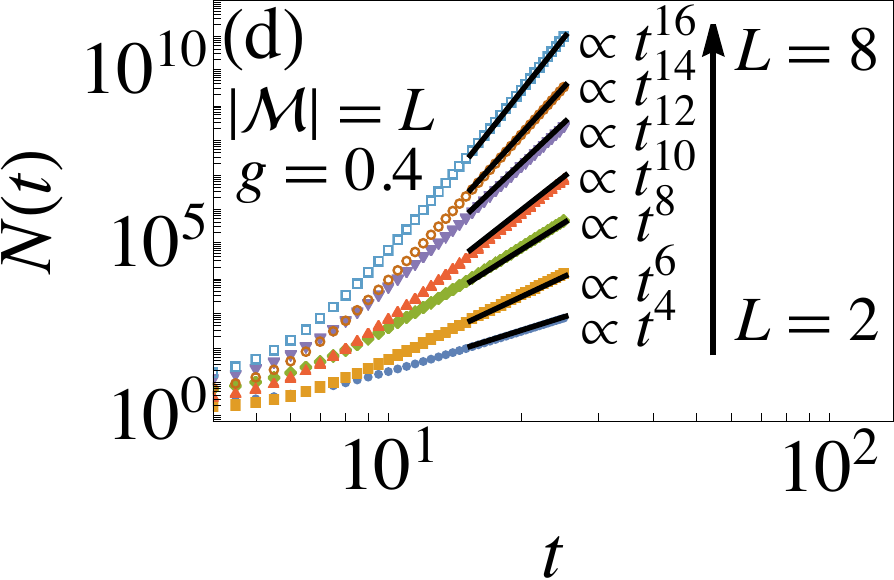}

\caption{\textcolor{red}{\label{fig:FHM_scaling_exp}}Quantum dynamics of Fermi
gases in optical lattices under monitoring. (a) Schematic illustration
of Fermi gases in optical lattices and the implementation of measurements.
(b) Time dependence of $N(t)$ at different $\gamma/g$ for $L=6$.
Power law scaling $N(t)\propto t^{12}$ emerges as $\gamma/g\rightarrow1$.
(c) Scalings of $N(t)$ at the critical point $\gamma/g=1$ for different
$(L,|\mathcal{M}|,g)$. (d) IFDS $N(t)\propto t^{2L}$ at the critical
point $\gamma/g=1$.  The simulations are performed with $J=1$, $U=5$,
$dt=10^{-3}$, and the initial state given by $|\psi(t=0)\rangle=|\uparrow_{1},\cdots,\uparrow_{L/2},\downarrow_{L/2+1},\cdots,\downarrow_{L}\rangle$,
with measurements conducted in the central region of the system.}
\end{figure}

\emph{IFDS in optical lattices.}\textemdash The preceding discussion
has focused on spin systems and we notice that the essential structure
that leads to the emergence of the IFDS is the symmetry that can be
assumed in the system. This motivates us to investigate the IFDS beyond
spin systems. One case in point is ultracold Fermi gases in optical
lattices, which not only respects the same $\mathrm{SU}(2)$ symmetry
as the Heisenberg spin chain, but is also highly pertinent to experiments
with ultracold atoms \citep{Robert2008nature,Michael2015science,Martin2016science,Mazurenko2017nature}.

As illustrated schematically in Fig.~\ref{fig:FHM_scaling_exp}(a),
the system being monitored is an ultracold spin-1/2 Fermi gas loaded
in a 1D optical lattice, a part of which is shined by a Rabi laser
that couples the two internal states of the atoms. An observer repeatedly
perform ``weak'' measurements on the occupation of spin-up atoms
in the shined region and decide whether the system continue evolving
according to the measurement outcome.

More specifically, the Hamiltonian of the system reads 
\begin{align}
\hat{H}_{\mathrm{F}}= & -J\sum_{j=1}^{L-1}\sum_{\sigma}(\hat{c}_{j,\sigma}^{\dagger}\hat{c}_{j+1,\sigma}+\mathrm{h}.\,\mathrm{c}.)+U\sum_{j=1}^{L}\hat{n}_{j,\uparrow}\hat{n}_{j,\downarrow}\label{eq:FHM_Rabi}\\
 & +g\sum_{j\in\mathcal{M}}(\hat{c}_{j,\uparrow}^{\dagger}\hat{c}_{j,\downarrow}+\hat{c}_{j,\downarrow}^{\dagger}\hat{c}_{j,\uparrow}),\nonumber 
\end{align}
where $\hat{c}_{j,\sigma}^{\dagger}$($\hat{c}_{j,\sigma}$) is the
creation (annihilation) operator of the fermions with spin $\sigma$
($\sigma=\uparrow,\downarrow$) at the $j$th site and $\hat{n}_{j,\sigma}\equiv\hat{c}_{j,\sigma}^{\dagger}\hat{c}_{j,\sigma}$
is the particle number operator counting the number of atoms with
spin $\sigma$ on site $j$. The first two terms in (\ref{eq:FHM_Rabi})
are the conventional hopping term and the on-site interaction term,
with their strengths being $J$ and $U$, respectively. The third
term is a spin flip term that describes the coupling between the Rabi
laser and the atoms within the region specified by $\mathcal{M}$,
with its strength being $g$. From the form of Hamiltonian (\ref{eq:FHM_Rabi}),
one can see that this system can be readily realized in current ultracold
atom experiments employing ultracold Fermi gases in optical lattices.

The measurement imposed on the system is described by the operator
$\hat{M}_{\mathrm{F}}=1-dt\gamma\sum_{j\in\mathcal{M}}(\hat{n}_{j,\uparrow}-\hat{n}_{j,\downarrow})$
\citep{Sup_Mat} with $\gamma$ characterizing the strength of the
measurement. For Fermi gases in deep optical lattices at half filling,
this corresponds to weakly measuring the occupation of spin-up atoms
in the region $\mathcal{M}$, followed by post-selection of the quantum
trajectories without spin-up atoms. Specifically, as schematically
illustrated in Fig.~\ref{fig:FHM_scaling_exp}(a), this measurement
process can be implemented in experiments by weakly coupling the internal
$|\uparrow\rangle$ state of the atoms to an auxiliary state $|a\rangle$
with a probe laser and subsequently performing postselction based
on the results of the strong measurements with respect to $|a\rangle$
\citep{Lee2014PRX,Sup_Mat}.

Fig.~\ref{fig:FHM_scaling_exp}(b) presents the time evolution of
the total particle number $N(t)\equiv\langle\psi(t)|\sum_{j,\sigma}\hat{n}_{j,\sigma}|\psi(t)\rangle$
at different measurement strengths for $L=6$, starting from an easily
accessible initial state in experiments, namely, $|\psi(0)\rangle=|\uparrow_{1},\cdots,\uparrow_{L/2},\downarrow_{L/2+1},\cdots,\downarrow_{L}\rangle$.
Notably, as the measurement strength $\gamma$ approaches the Rabi
coupling strength $g$, $N(t)$ exhibits a clear dynamical scaling
$\propto t^{12}$. To explore the possible IFDS of the system, we
further calculate $N(t)$ of the systems at different $(L,|\mathcal{M}|,g)$
as shown in Figs.~\ref{fig:FHM_scaling_exp}(c,~d). From these results,
one notices that an IFDS, $N(t)\propto t^{2L}$, indeed exists for
the system, regardless of the size of the region under monitoring.
This clearly suggests the robustness of the observed IFDS, making
them readily accessible for experimental observation. In particular,
for current ultracold atom experiments in the deep optical lattices
and half-filling regime, the measurement protocol can be relatively
easily implemented as discussed above (see SM \citep{Sup_Mat} for
more general cases without deep optical lattices and more details
on the experimental observability).

\emph{Conclusion and discussion}.\textemdash We have revealed that
IFDS can emerge in quantum many-body systems under monitoring, with
the scaling exponents determined by the order of the hierarchy of
EPs present in the system's effective non-Hermitian description. Although
our investigation has focused on 1D systems, it is noteworthy that
IFDS can also exist in high-dimensional Heisenberg spin systems and
Fermi gases in optical lattices as the key mechanism underlying this
IFDS is the internal $\mathrm{SU}(2)$ symmetry. Moreover, the thermodynamic
limit is not mandatory for the manifestation of these distinctive
dynamical scaling behavior, greatly facilitating the direct experimental
observation of IFDS across a plethora of platforms, such as ultracold
atoms in optical lattices \citep{Bloch_RMP_2008,Robert2008nature,Michael2015science,Martin2016science,Mazurenko2017nature,Zhang2018AP,Florian2020NRP},
the trapped-ion \citep{Monroe2021RMP} or superconducting quantum
computing systems \citep{georgescu2014RMP}. Our predictions will
also stimulate further research on the critical scaling in measurement
induced transitions, encompassing diverse measurement protocols like
random projective selections.
\begin{acknowledgments}
We thank Guo-Qing Zhang and Zheng-Yuan Xue for helpful discussions.
This work was supported by NKRDPC (Grant~Nos.~2022YFA1405300), NSFC
(Grant Nos.~12074180, and 12275089), Guangdong Basic and Applied
Research Foundation (Grant~No.~2023A1515012800), Guangdong Provincial
Key Laboratory (Grant No. 2020B1212060066), Key-Area Research and
Development Program of Guangdong Province (Grant No. 2019B030330001),
and START grant of South China Normal University. 
\end{acknowledgments}

\bibliographystyle{apsrev4-1}

%merlin.mbs apsrev4-1.bst 2010-07-25 4.21a (PWD, AO, DPC) hacked
%Control: key (0)
%Control: author (72) initials jnrlst
%Control: editor formatted (1) identically to author
%Control: production of article title (-1) disabled
%Control: page (0) single
%Control: year (1) truncated
%Control: production of eprint (0) enabled
%

%%%%%%%%%% Merge with supplemental materials %%%%%%%%%%
\clearpage{}
\begin{widetext}
\begin{center}
\textbf{\large Supplemental Material for ``Measurement-induced integer families
of critical dynamical scaling in quantum many-body systems''}
\end{center}
\end{widetext}
%%%%%%%%%% Merge with supplemental materials %%%%%%%%%%
%%%%%%%%%% Prefix a "S" to all equations, figures, tables and reset the counter %%%%%%%%%%
\setcounter{equation}{0}
\setcounter{figure}{0}
\setcounter{table}{0}
\setcounter{page}{1}
\makeatletter
\renewcommand{\theequation}{S\arabic{equation}}
\renewcommand{\thefigure}{S\arabic{figure}}
%%%%%%%%%% Prefix a "S" to all equations, figures, tables and reset the counter %%%%%%%%%%
\section{Measurement Protocol}

The quantum many-body systems we discuss in the main text undergo
measurements and post-selections after a brief period of unitary evolution
(see Fig.~1). These processes are facilitated by the operator $\hat{M}$,
which varies in form for spin ($\hat{M}_{\text{S}}$) and Fermi gases
($\hat{M}_{\text{F}}$) systems. This section elaborates on the experimental
implementation of these measurement operators.

\subsection{Single-site Measurement}

We begin by discussing single-site measurement, corresponding to $|\mathcal{M}|=1$.
Without loss of generality, we consider the measurement and post-selection
operator realized by $\hat{M}_{\mathrm{Exp}}=1-k\hat{n}$, where $\hat{n}$
has eigenvalues $1$ and $0$ for eigenstates $|1\rangle$ and $|0\rangle$
respectively. The relationship between $\hat{M}_{\mathrm{Exp}}$ and
$\hat{M}_{\text{S}},\hat{M}_{\text{F}}$ will be clarified after general
discussions.

Consider a general local state after unitary evolution and before
measurement, which can be expressed as 
\begin{equation}
|\psi(t)\rangle=c_{1}|1\rangle+c_{0}|0\rangle.
\end{equation}
We couple $|1\rangle$ to an auxiliary state, $|a\rangle$, with a
transition probability $p$ (see Fig.~3a). The state after coupling
is 
\begin{equation}
c_{1}\sqrt{p}|a\rangle+c_{1}\sqrt{1-p}|1\rangle+c_{0}|0\rangle.
\end{equation}
We can then design a scheme to detect whether the atom is in the auxiliary
state, like the protocol using the signal of fluorescence in \citep{Lee2014PRX}.
If the auxiliary state is detected, we discard this trajectory. The
state in the remaining trajectories, in the absence of an auxiliary
state under a small-$p$ limit, takes the form, 
\begin{align}
|\psi(t+dt)\rangle & =c_{1}(1-\frac{p}{2})|1\rangle+c_{0}|0\rangle\nonumber \\
 & =(1-\frac{p}{2}\hat{n})|\psi(t)\rangle.
\end{align}
It can be seen that the above post-selected state exactly coincides
with $\hat{M}_{\mathrm{Exp}}|\psi(t)\rangle$ under the condition
$p=2k$.

\subsection{Many-site Measurement}

To extend to many-site measurement cases, we first consider the two-site
measurement case, where the measurement operator we aim to realize
is $\hat{M}_{\mathrm{Exp}}=1-k(\hat{n}_{1}+\hat{n}_{2})$. The subscript
in $\hat{n}_{j}$ for $j=1,2$ denotes the locations. Given a general
initial state 
\begin{equation}
|\psi(t)\rangle=c_{11}|11\rangle+c_{10}|10\rangle+c_{01}|01\rangle+c_{00}|00\rangle,
\end{equation}
we locally couple the two $|1\rangle$ states to two auxiliary states
$|a_{1}\rangle$ and $|a_{2}\rangle$, respectively. The state after
coupling is expressed as 
\begin{align}
c_{11} & \left[(\sqrt{p})^{2}|a_{1}a_{2}\rangle+\sqrt{p}\sqrt{1-p}|a_{1}1\rangle\right.\nonumber \\
 & \left.+\sqrt{1-p}\sqrt{p}|1a_{2}\rangle+(\sqrt{1-p})^{2}|11\rangle\right]\nonumber \\
+c_{10} & \left[\sqrt{p}|a_{1}0\rangle+\sqrt{1-p}|10\rangle\right]\nonumber \\
+c_{01} & \left[\sqrt{p}|0a_{2}\rangle+\sqrt{1-p}|01\rangle\right]\nonumber \\
+c_{00} & |00\rangle.
\end{align}
As before, we discard all states in the presence of $|a_{1}\rangle$
or $|a_{2}\rangle$ such that the state in the trajectories with the
absence of an auxiliary state under a small-$p$ limit is given by
\begin{align}
|\psi(t+dt)\rangle= & c_{11}(1-p)|11\rangle+c_{10}(1-\frac{p}{2})|10\rangle\nonumber \\
 & +c_{01}(1-\frac{p}{2})|01\rangle+c_{00}|00\rangle\nonumber \\
= & (1-\frac{p}{2}\hat{n}_{1}-\frac{p}{2}\hat{n}_{2})|\psi(t)\rangle.
\end{align}
This state exactly coincides with $\hat{M}_{\mathrm{Exp}}|\psi(t)\rangle$
under the same condition $p=2k$.

Therefore, the measurement protocol for many-site measurement cases
with an operator $\hat{M}_{\mathrm{Exp}}=1-k\sum_{j\in\mathcal{M}}\hat{n}_{j}$
can be straightforwardly generalized from the above two-site cases.
In short, given a general initial state, we locally couple all the
$|1\rangle$ states to a series of auxiliary states with a probability
$p=2k$ of transitioning to it. The relative probability of a state
that does not couple to the auxiliary state is $(\sqrt{1-2k})^{\langle\sum_{j\in\mathcal{M}}\hat{n}_{j}\rangle}$,
where $\langle\sum_{j\in\mathcal{M}}\hat{n}_{j}\rangle$ counts the
total number of the local $|1\rangle$ states within the measurement
region $\mathcal{M}$. After post-selection, the final state coincides
with the action of $\hat{M}_{\mathrm{Exp}}$ on the initial state
in the small-$k$ limit.

\subsection{Fermi Gases}

For Fermi gases in deep optical lattices at half filling, the measurement
operator in the small-$dt$ limit is approximately 
\begin{align}
\hat{M}_{\mathrm{F}} & \approx1-dt\gamma\sum_{j\in\mathcal{M}}2\hat{n}_{j,\uparrow}+dt\gamma|\mathcal{M}|\nonumber \\
 & \approx e^{dt\gamma|\mathcal{M}|}(1-2dt\gamma\sum_{j\in\mathcal{M}}\hat{n}_{j,\uparrow}).\label{eq:Simple_M_F}
\end{align}
For a generic state in the large-$U$ limit, $\left|\psi(t)\right\rangle =\sum_{\{\sigma_{j}\}}c_{\{\sigma_{j}\}}|\sigma_{1}\sigma_{2}\cdots\sigma_{L}\rangle$
($\sigma=\uparrow,\downarrow$), the state after acting the measurement
operator reads 
\begin{align}
 & \hat{M}_{\mathrm{F}}\left|\psi(t)\right\rangle \nonumber \\
\approx & e^{dt\gamma|\mathcal{M}|}(1-2dt\gamma\sum_{j\in\mathcal{M}}\hat{n}_{j,\uparrow})\left|\psi(t)\right\rangle .
\end{align}
This measurement operator can be realized by the above $\hat{M}_{\mathrm{Exp}}=1-k\sum_{j\in\mathcal{M}}\hat{n}_{j}$
accompanied by a decay factor $e^{dt\gamma|\mathcal{M}|}$ once we
make the following identifications 
\begin{align}
k & =2dt\gamma,\\
|0\rangle & =|\downarrow\rangle,\,|1\rangle=|\uparrow\rangle,\\
\hat{n}_{j} & =\hat{n}_{j,\uparrow}.
\end{align}

\subsection{Spin Model}

For the spin model, the measurement operator takes the form \textcolor{black}{
\begin{align}
\hat{M}_{\text{S}} & =1-\gamma dt\sum_{j\in\mathcal{M}}(1-\hat{\sigma}_{j}^{y}).
\end{align}
}It can be realized by $\hat{M}_{\mathrm{Exp}}=1-k\sum_{j\in\mathcal{M}}\hat{n}_{j}$
if we make the identifications, 
\begin{align}
k & =2dt\gamma,\\
\hat{\sigma}^{y}|0\rangle & =+1|0\rangle,\,\hat{\sigma}^{y}|1\rangle=-1|1\rangle,\\
\hat{n}_{j} & =(1-\hat{\sigma}_{j}^{y})/2.
\end{align}

\section{Further Insights into Integer Families of Dynamical Scaling}

In the main text, we demonstrated that the late-time dynamics of certain
observables (such as total magnetization in spin chains and total
particle number in Fermi gases) exhibit power-law scaling. The exponents
of this scaling linearly depend on the size of the measurement region
or the lattice of the system. In this section, we provide additional
numerical evidence for the integer family of dynamical scaling in
transverse Ising and Heisenberg spin models. We also illustrate how
the initial state influences the onset of dynamical scaling in Fermi
gases in optical lattices.

\subsection{Heisenberg chain}

\begin{figure*}
\includegraphics[width=2.2in]{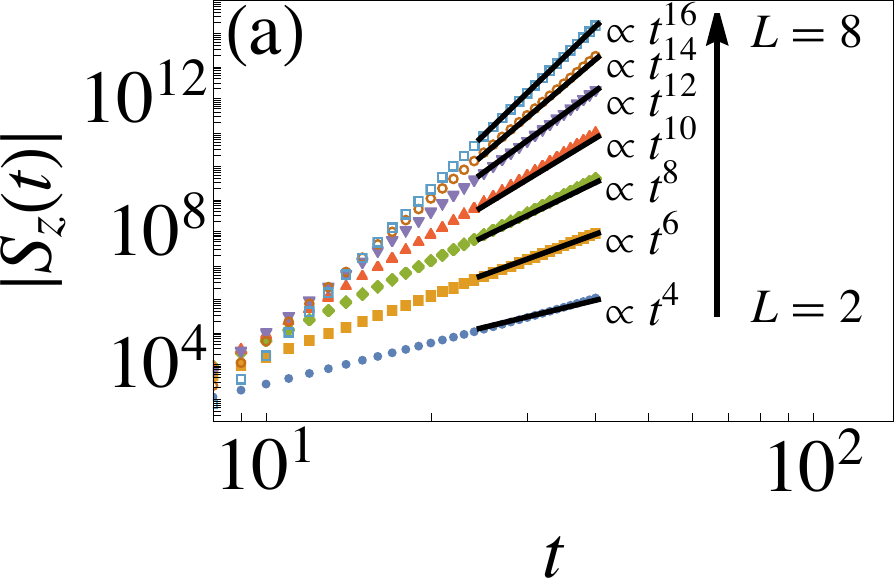}~\includegraphics[width=2.2in]{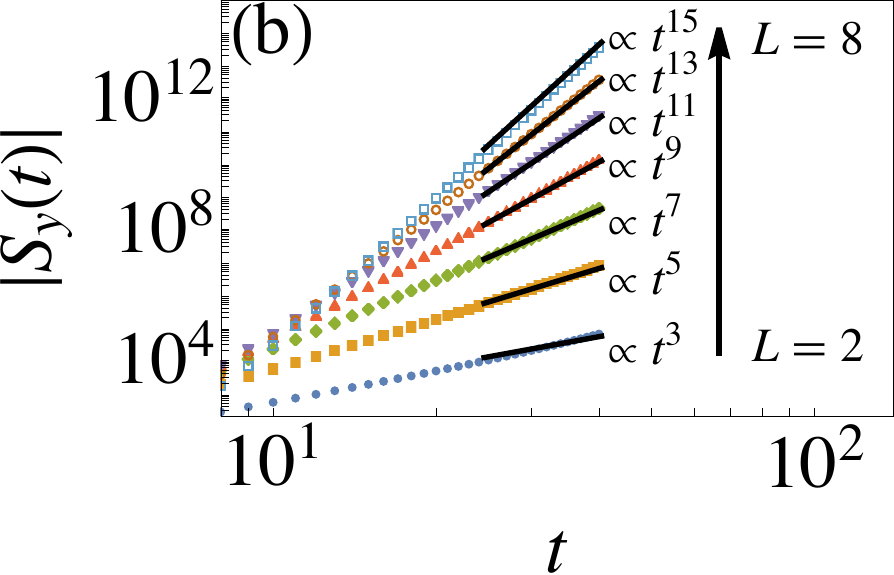}~\includegraphics[width=2.2in]{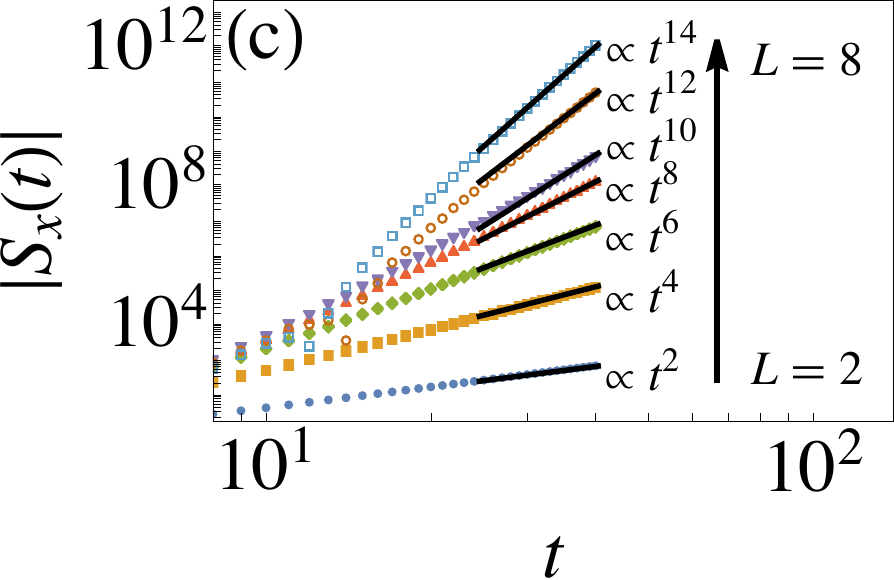}\caption{\label{fig:Heisenberg_dynamics}Quantum dynamics of the transverse
Heisenberg model under monitoring. All dynamics were simulated with
$J=g=\gamma=1$, $dt=10^{-3}$, $\mathcal{M}=\{1\}$, and the initial
state given by $|\psi(t=0)\rangle=2^{-L}\sum_{\{\sigma_{i}\}}|\sigma_{1}\sigma_{2}\cdots\sigma_{L}\rangle$.
(a-c) Power law scaling of $|S_{z}(t)|$ (a), $|S_{y}(t)|$ (b), and
$|S_{x}(t)|$ (c) for different system sizes. The late-time dynamics
of $|S_{z}(t)|$ (a), $|S_{y}(t)|$ (b), and $|S_{x}(t)|$ (c) manifest
the integer family of scaling $|S_{z}(t)|\propto t^{2L}$, $|S_{y}(t)|\propto t^{2L-1}$,
and $|S_{x}(t)|\propto t^{2L-2}$, respectively. See the supplementary
text for more details.}
\end{figure*}

Figure \ref{fig:Heisenberg_dynamics} illustrates the time evolution
of $|S_{z}(t)|$, $|S_{y}(t)|$, and $|S_{x}(t)|$ in the transverse
Ising model under monitoring for different system sizes $L$. For
$|S_{z}(t)|$ in Fig.~\ref{fig:Heisenberg_dynamics}(a), we find
that the late-time dynamical scaling can be fitted by the empirical
formula $|S_{z}(t)|\propto t^{2L}$. Furthermore, the integer family
for other observables, such as the total magnetization along other
directions, suggests that the integer family of dynamical scaling
takes the form $t^{2L-q}$ for some non-negative integer $q$. For
instance, we observe that $q=1$ for $|S_{y}(t)|$ in Fig.~\ref{fig:Heisenberg_dynamics}(b)
and $q=2$ for $|S_{x}(t)|$ in Fig.~\ref{fig:Heisenberg_dynamics}(c).

\subsection{Ising chain}

\begin{figure*}
\includegraphics[width=2.2in]{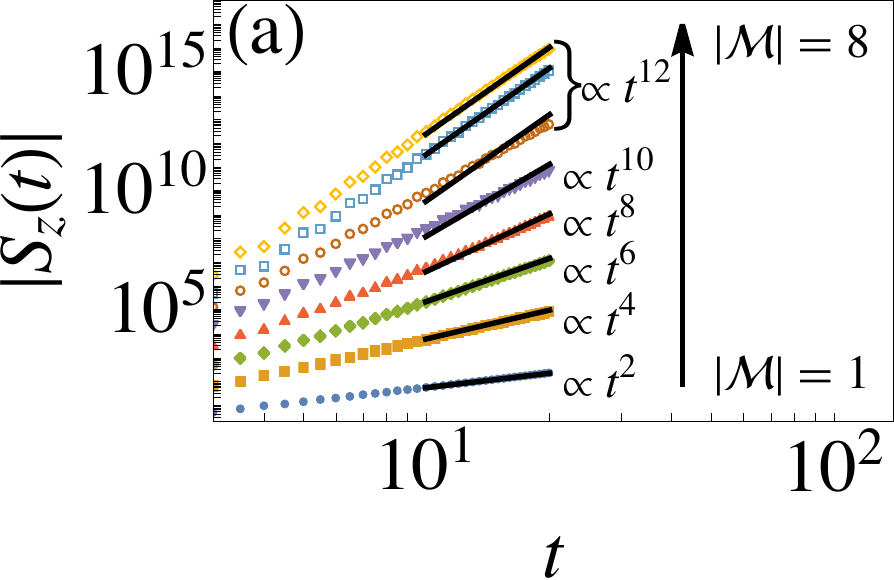}~\includegraphics[width=2.2in]{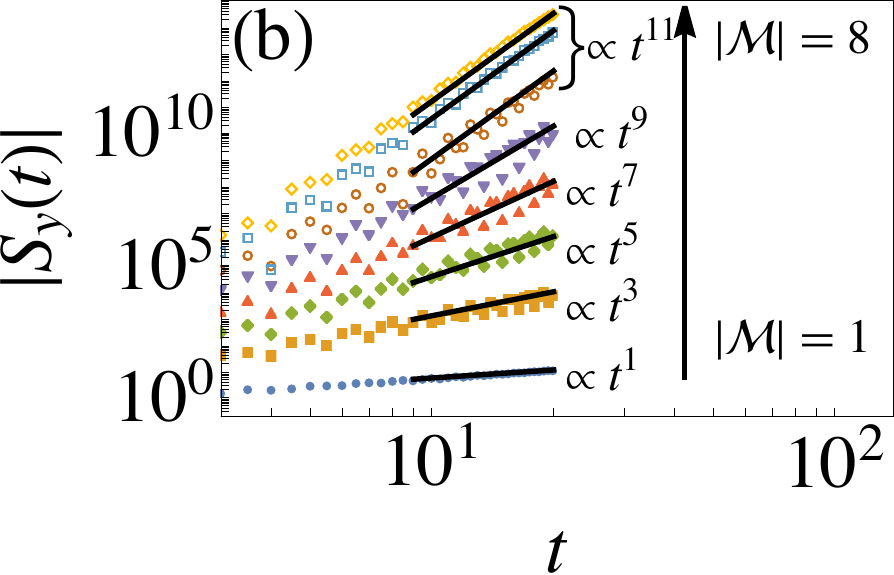}~\includegraphics[width=2.2in]{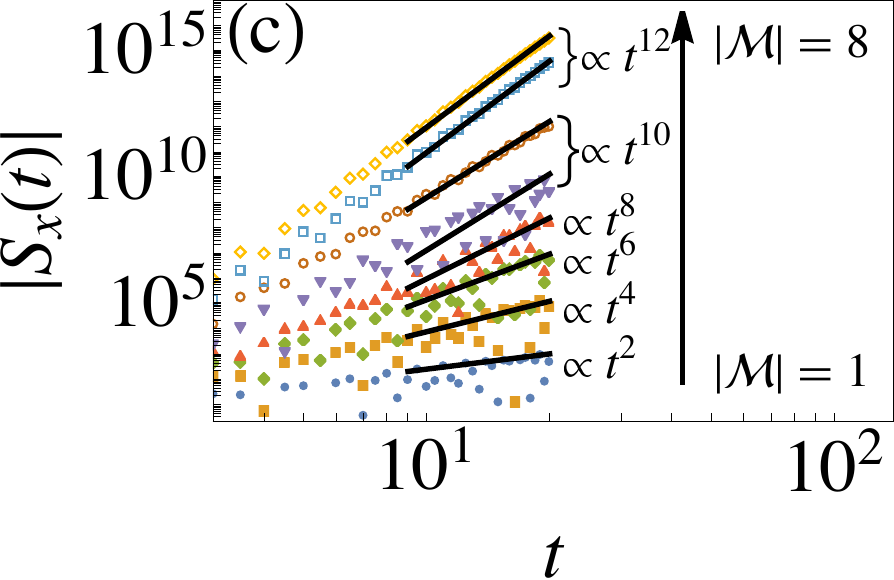}

\caption{\label{fig:Ising_dynamics}Quantum dynamics of the transverse Ising
model under monitoring. All dynamics were simulated with $J=g=\gamma=1$,
$dt=10^{-4}$, $L=8$, $\mathcal{M}=\{1,2,\cdots,|\mathcal{M}|\}$,
and the initial state given by $|\psi(t=0)\rangle=2^{-L}\sum_{\{\sigma_{i}\}}|\sigma_{1}\sigma_{2}\cdots\sigma_{L}\rangle$.
(a-c) Power law scaling of $|S_{z}(t)|$ (a), $|S_{y}(t)|$ (b), and
$|S_{x}(t)|$ (c) for different sizes of measurement regions. The
late-time dynamics of $|S_{z}(t)|$ (a) manifests an integer family
of scaling $|S_{z}(t)|\propto t^{2|\mathcal{M}|}$ for $|\mathcal{M}|\protect\leq L-2$.
The late-time dynamics of $|S_{y}(t)|$ (b) and $|S_{x}(t)|$ (c)
exhibit the integer family of scaling $|S_{y}(t)|\propto t^{2|\mathcal{M}|-1}$
for $|\mathcal{M}|\protect\leq L-2$ and $|S_{x}(t)|\propto t^{2|\mathcal{M}|}$
for $|\mathcal{M}|\protect\leq L-3$, respectively. See the supplementary
text for more details.}
\end{figure*}

Figure \ref{fig:Ising_dynamics} illustrates the time evolution of
$|S_{z}(t)|$, $|S_{y}(t)|$, and $|S_{x}(t)|$ in the transverse
Ising model under monitoring for different sizes of measurement regions
$\mathcal{M}$. For $|S_{z}(t)|$ in Fig.~\ref{fig:Ising_dynamics}(a),
we find that the late-time dynamical scaling can be fitted by the
empirical formula $|S_{z}(t)|\propto t^{2|\mathcal{M}|}$ for $|\mathcal{M}|\leq L-2$.
Furthermore, the integer family for other observables, such as the
total magnetization along other directions, suggests that the integer
family of dynamical scaling takes the form $t^{2|\mathcal{M}|-q}$
within a certain valid region of $\mathcal{M}$ for some non-negative
integer $q$. For instance, we observe that $q=1$ for $|S_{y}(t)|$
in Fig.~\ref{fig:Ising_dynamics}(b) (valid for $|\mathcal{M}|\leq L-2$)
and $q=0$ for $|S_{x}(t)|$ in Fig.~\ref{fig:Ising_dynamics}(c)
(valid for $|\mathcal{M}|\leq L-3$).

\subsection{Fermi gases in optical lattices}

\begin{figure}
\includegraphics[width=2in]{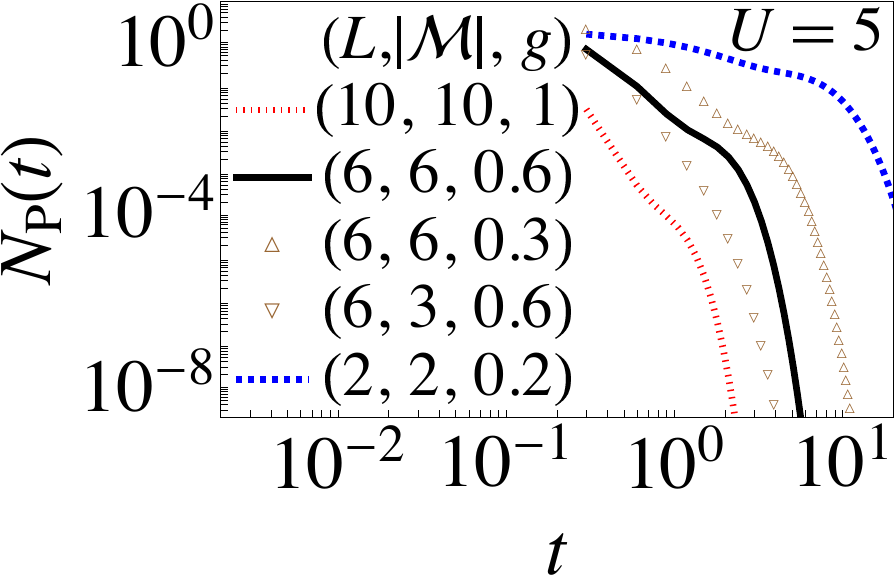}

\caption{\label{fig:Fermi_Gases_largeU_remaining}Time dependence of the remaining
particle number $N_{\mathrm{P}}$ in Fig.~3(c) after post-selections,
which is measured in experiments. The simulations were performed with
$J=1$, $dt=10^{-3}$, and the initial state given by $|\psi(t=0)\rangle=|\uparrow_{1},\cdots,\uparrow_{L/2},\downarrow_{L/2+1},\cdots,\downarrow_{L}\rangle$,
with measurements conducted in the central region of the system. See
the supplementary text for more details. }
\end{figure}

To assess the experimental complexity in this case of Fig.~3(c),
we further investigate the remaining particle number $N_{\mathrm{P}}$
after postselections that is actually measured in experiments. Our
analysis indicates that, after continuous discarding the experiment
outputs, for system sizes $L=2,\,6$, and $10$, one needs to run
about $10^{1}$, $10^{3}$, and $10^{4}$ times of experiments to
construct $N_{\mathrm{P}}(t)$ in practice, respectively (see Fig.~\ref{fig:Fermi_Gases_largeU_remaining}).
This suggests that direct measurement of these scaling is indeed achievable
with current experimental technology. 

\begin{figure}
\includegraphics[width=1.7in]{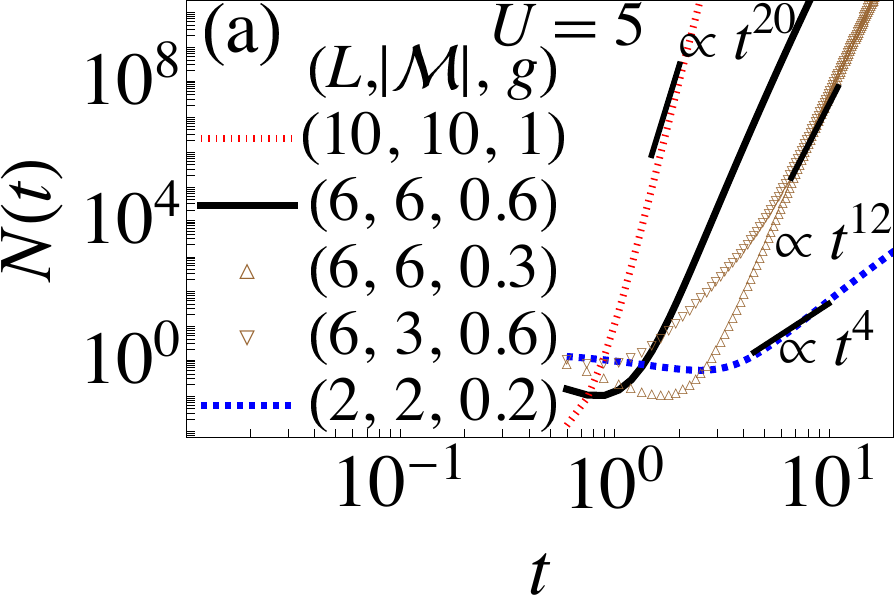}~\includegraphics[width=1.7in]{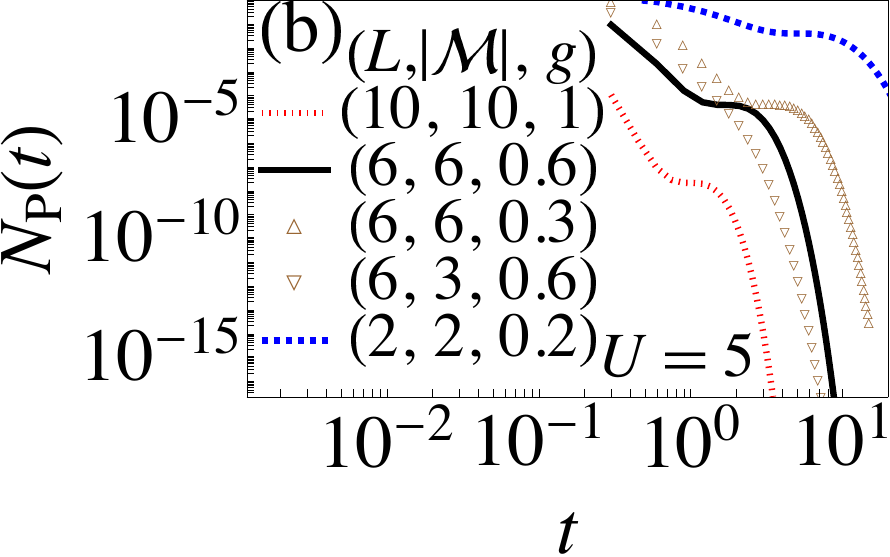}

\caption{\label{fig:Fermi_Gases_dynamics}Quantum dynamics of Fermi gases in
optical lattices under monitoring. The simulations were performed
with $J=1$, $U=5$, $dt=10^{-3}$, and the initial state given by
$|\psi(t=0)\rangle=|\uparrow_{1},\uparrow_{2},\cdots,\uparrow_{L}\rangle$,
with measurements conducted in the central region of the system. (a)
Power law scaling of $N(t)$ for various system and measurement region
sizes exhibits an integer family of scaling $N(t)\propto t^{2L}$.
(b) Time dependence of the remaining particle number $N_{\mathrm{P}}$
after post-selections, which is measured in experiments. See the supplementary
text for more details.}
\end{figure}

We also performed numerical simulations for the Fermi gases model
in the large-$U$ limit, using the initial state $|\uparrow_{1},\uparrow_{2},\cdots,\uparrow_{L}\rangle$
(Fig.~\ref{fig:Fermi_Gases_dynamics}). All other conditions were
the same as those for the initial state $|\uparrow_{1},\cdots,\uparrow_{L/2},\downarrow_{L/2+1},\cdots,\downarrow_{L}\rangle$
in the main text. As shown in Fig.~\ref{fig:Fermi_Gases_dynamics}(a),
the late-time behaviors for the all-polarized initial state are the
same as those for the half-polarized initial state, i.e., the total
particle number exhibits the integer family of dynamical scaling $N(t)\propto t^{2L}$.
The main difference can be seen in Fig.~\ref{fig:Fermi_Gases_dynamics}(b),
which indicates that after continuously discarding the experimental
outputs, for system sizes $L=2$, $6$, and $10$, one needs to run
about $10^{2}$, $10^{4}$, and $10^{6}$ times of experiments to
construct $N_{\text{P}}(t)$ in practice, respectively.

\begin{figure}
\includegraphics[width=1.7in]{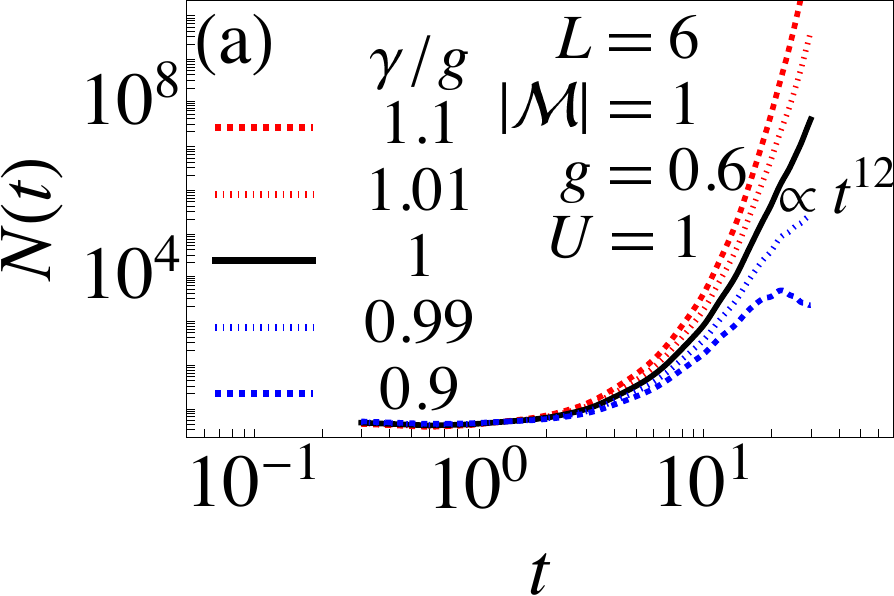}~\includegraphics[width=1.7in]{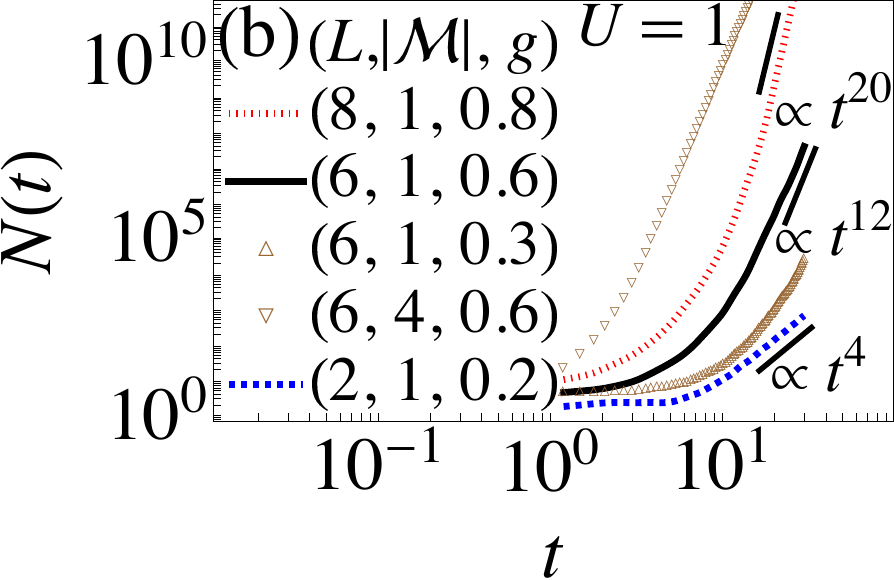}

\caption{\label{fig:Fermi_Gases_small_U}Quantum dynamics of Fermi gases in
optical lattices under monitoring. The simulations were performed
with $J=1$, $U=1$, $dt=10^{-3}$, and the initial state given by
$|\psi(t=0)\rangle=|\uparrow_{1},\cdots,\uparrow_{L/2},\downarrow_{L/2+1},\cdots,\downarrow_{L}\rangle$,
with measurements conducted in the central region of the system. (a)
Time dependance of $N(t)$ at different $\gamma/g$ for $L=6$. Power
law scaling $N(t)\propto t^{12}$ emerges as $\gamma/g\rightarrow1$.
(b) Scalings of $N(t)$ at the critical point $\gamma/g=1$ for intermediate
on-site interaction strengths. See the supplementary text for more
details.}
\end{figure}
For more general cases without a deep optical lattice, the measurement
operator is 
\begin{equation}
\hat{M}_{\mathrm{F}}=1-dt\gamma\sum_{j\in\mathcal{M}}(\hat{n}_{j,\uparrow}-\hat{n}_{j,\downarrow}).
\end{equation}
This original form of measurement operator is more complicated than
its simplified version in the large-$U$ limit in Eq.~(\ref{eq:Simple_M_F}).
As shown in Figs.~\ref{fig:Fermi_Gases_small_U}(a,~b), all the
scaling behaviors are the same as those in the deep optical lattice,
where the power law scaling $N(t)\propto t^{12}$ emerges as $\gamma/g\rightarrow1$. 

\section{\label{sec:Higher-order-EPs}Higher Order EPs in effective Hamiltonians
of Measurement Dynamics}

Generally regarding measurements as influences imposed by the environment,
the dynamics discussed in this work fall under the category of open
quantum systems, typically described by the Lindblad master equation
\citep{Gardiner_2000_Springer,Breuer_2000_Oxford}. However, in contrast
to conventional environments where the influences are essentially
random at each instant, in this scenario, the environment's effects
are explicitly known through the measurement results. Consequently,
this significantly simplifies the effective description of the dynamics
and makes it possible to construct an effective non-Hermitian Hamiltonian
\citep{gopalakrishnan_Gullans2021PRL,Turkeshi2021PRB,Jian2021PRB}
that faithfully captures the system's physics in the presence of measurements.
This stands in stark contrast to the conventional open quantum systems
where non-Hermitian effective Hamiltonians typically emerge as approximate
descriptions to replace Lindblad master equations \citep{Ashida2020AP}.
Here, these effective non-Hermitian Hamiltonians are instead promoted
as faithful descriptions of the system's dynamics.

In the limit of continuous measurement, the effective Hamiltonians
for the measurement dynamics in Ising ($\hat{H}_{\mathrm{TI}}^{\mathrm{eff}}$)
and Heisenberg ($\hat{H}_{\mathrm{TH}}^{\mathrm{eff}}$) chains are
found to contain the following non-Hermitian term 
\begin{equation}
\sum_{j\in\mathcal{M}}[g\hat{\sigma}_{j}^{x}+i\gamma(1-\hat{\sigma}_{j}^{y})].
\end{equation}
Upon first inspection of this non-Hermitian term at $g=\gamma$, we
discern that the local components, $\hat{\sigma}_{j}^{x}-i\hat{\sigma}_{j}^{y}$,
support an EP of order two. However, the EPs of individual components
do not guarantee the presence of an EP in the complete Hamiltonian.
Recognizing that any arbitrary matrix can be transformed into its
Jordan normal form, which is block diagonalized with the Jordan sub-block
\begin{equation}
J_{n}(\lambda)=\lambda I_{n}+J_{n}^{+},
\end{equation}
where $I_{n}$ is the $n\times n$ identity matrix, and 
\begin{equation}
J_{n}^{+}\equiv\left(\begin{array}{cccccc}
0 & 1 & 0 & \cdots & 0 & 0\\
0 & 0 & 1 & \cdots & 0 & 0\\
\cdots & \cdots & \cdots & \cdots & \cdots & \cdots\\
0 & 0 & 0 & \cdots & 0 & 1\\
0 & 0 & 0 & \cdots & 0 & 0
\end{array}\right),
\end{equation}
we can see that if the matrix is diagonalizable, the Jordan sub-blocks
have size $1$, i.e., $n=1$. For a Hamiltonian operator containing
an EP of order $N$, its matrix representation is nondiagonalizable
and thus possesses a Jordan block of size $N$. To ascertain the existence
and the order of EPs in the effective Hamiltonian of transverse Heisenberg
and Ising models under measurements, we should identify its Jordan
normal form, from which the order of the EP can be directly gleaned
based on the size of the Jordan sub-blocks.

\subsection{Heisenberg Chain}

\subsubsection{Global Defective Term}

To facilitate our discussion, we operate with the spin operators $\hat{S}^{x(y)(z)}\equiv\frac{1}{2}\hat{\sigma}^{x(y)(z)}$
($\hbar=1$), and redefine the energy scale $J$ and $g$ so that
the effective Hamiltonian at $\gamma=g$ is expressed as 
\begin{equation}
\hat{H}=-J\sum_{j=1}^{L-1}\hat{\mathbf{S}}_{j}\cdot\hat{\mathbf{S}}_{j+1}-g\sum_{j\in\mathcal{M}}\hat{S}_{j}^{-},
\end{equation}
where $\sum_{j=1}^{L-1}\hat{\mathbf{S}}_{j}\cdot\hat{\mathbf{S}}_{j+1}\equiv\sum_{a=x,y,z}\hat{S}_{j}^{a}\hat{S}_{j+1}^{a}$,
$\hat{S}^{\pm}\equiv\hat{S}^{x}\pm i\hat{S}^{y}$, and a trivial constant
$-i\gamma|\mathcal{M}|$ is disregarded. In the following, we will
consider the simplest case with an open boundary condition and a global
transverse field under global measurement, i.e., $\mathcal{M}=\{1,2,\cdots,L\}$.

Notice that the interaction term $-J\sum_{j=1}^{L-1}\hat{\mathbf{S}}_{j}\cdot\hat{\mathbf{S}}_{j+1}$
commutes with the three total spin operators ($\hat{S}^{x(y)(z)}\equiv\sum_{j=1}^{L}\hat{S}_{j}^{x(y)(z)}$).
This term is invariant with respect to total spin rotation, which
is the origin of saying the interaction term exhibiting a $\text{SU}(2)$
symmetry. This symmetry has two significant implications. Firstly,
the interaction term commutes with $\hat{S}^{2}$ and $\hat{S}^{z}$.
As a result, the set $\{-J\sum_{j=1}^{L-1}\hat{\mathbf{S}}_{j}\cdot\hat{\mathbf{S}}_{j+1},\hat{S}^{2},\hat{S}^{z}\}$,
due to the relation $[\hat{S}^{2},\hat{S}^{z}]=0$, shares the same
eigenstates, denoted as $\{|n,s,m_{s}\rangle\}$. More specifically,
we have 
\begin{align}
-J\sum_{j=1}^{L-1}\hat{\mathbf{S}}_{j}\cdot\hat{\mathbf{S}}_{j+1}|n,s,m_{s}\rangle & =E_{n}|n,s,m_{s}\rangle,\\
\hat{S}^{2}|n,s,m_{s}\rangle & =(s+1)s|n,s,m_{s}\rangle,\\
\hat{S}^{z}|n,s,m_{s}\rangle & =m_{s}|n,s,m_{s}\rangle.
\end{align}
Secondly, the finite rotation operator $\mathscr{D}(\hat{\mathbf{n}},\theta)\equiv\exp(-i\hat{\mathbf{n}}\cdot\hat{\mathbf{S}}\theta)$
commutes with $-J\sum_{j=1}^{L-1}\hat{\mathbf{S}}_{j}\cdot\hat{\mathbf{S}}_{j+1}$.
This fact highlights the degeneracy of each eigenenergy $E_{n}$,
expressed as 
\begin{align}
-J\sum_{j=1}^{L-1}\hat{\mathbf{S}}_{j}\cdot\hat{\mathbf{S}}_{j+1} & \left(\mathscr{D}(\hat{\mathbf{n}},\theta)|n,s,m_{s}\rangle\right)\nonumber \\
=E_{n} & \left(\mathscr{D}(\hat{\mathbf{n}},\theta)|n,s,m_{s}\rangle\right).
\end{align}
The action of $\mathscr{D}(\hat{\mathbf{n}},\theta)$ on $|n,s,m_{s}\rangle$
intermixes all $|n,s,m_{s}\rangle$ states for $m_{s}=-s,\ldots,s$.
Therefore, each state $|n,s,m_{s}\rangle$ corresponds to the same
energy $E_{n}$ for all $m_{s}=-s,\ldots s$. This relationship can
be understood intuitively by recognizing that the ladder operator
commutes with the spin-interaction term, i.e., $[\hat{S}^{\pm},-J\sum_{j=1}^{L-1}\hat{\mathbf{S}}_{j}\cdot\hat{\mathbf{S}}_{j+1}]=0$.
Consequently, we find that $\hat{S}^{\pm}|n,s,m_{s}\rangle=\sqrt{(s\mp m_{s})(s\pm m_{s}+1)}|n,s,m_{s}\pm1\rangle$
possesses the same eigenenergy ($E_{n}$) as $|n,s,m_{s}\rangle$.

Considering the operator $\hat{S}^{\pm}$ ($\equiv\sum_{j=1}^{L}\hat{S}_{j}^{\pm}$),
it exhibits simple matrix elements within the total spin representation.
Leveraging the basic properties of ladder operators, we obtain 
\begin{align}
 & \langle s^{\prime},m_{s^{\prime}}^{\prime}|\hat{S}^{\pm}|s,m_{s}\rangle\nonumber \\
= & \sqrt{(s\mp m_{s})(s\pm m_{s}+1)}\delta_{s^{\prime},s}\delta_{m_{s^{\prime}}^{\prime},m_{s}\pm1},
\end{align}
from which we infer that the matrix of $\hat{S}^{\pm}$ is similar
to a Jordan normal matrix, up to the factor $\sqrt{(s\mp m)(s\pm m+1)}$.
This implies that it is block-diagonalized according to different
total spins and it is similar to a Jordan sub-block within each degenerate
subspace.

From the discussions above, we conclude that in the representation
related to the basis $\{|n,s_{n},m_{s}\rangle\}$ (where $s_{n}$
denotes the value of $s$ in state $|n\rangle$), the spin interaction
term $-J\sum_{j=1}^{L-1}\hat{\mathbf{S}}_{j}\cdot\hat{\mathbf{S}}_{j+1}$
takes the form 
\begin{align}
\stackrel[n=0]{n_{\max}}{\oplus}E_{n}I_{2s_{n}+1} & =\left(\begin{array}{ccc}
\ddots\\
 & \overset{(2s_{n}+1)\times(2s_{n}+1)}{\overbrace{\left(\begin{array}{ccccc}
E_{n} & 0 & 0 & \cdots & 0\\
0 & E_{n} & 0 & \cdots & \vdots\\
\vdots & 0 & \ddots & \cdots & \vdots\\
0 & \cdots & 0 & E_{n} & 0\\
0 & \cdots & \cdots & 0 & E_{n}
\end{array}\right)}}\\
 &  & \ddots
\end{array}\right)
\end{align}
and the defective part, $-g\hat{S}^{-}$, takes the form 
\begin{align}
\stackrel[n=0]{n_{\max}}{\oplus}\mathbb{J}_{2s_{n}+1} & =\left(\begin{array}{ccc}
\ddots\\
 & \overset{(2s_{n}+1)\times(2s_{n}+1)}{\overbrace{\left(\begin{array}{cccc}
0 & -g\sqrt{2s} & \cdots & 0\\
\vdots & 0 & \ddots & \vdots\\
0 & \cdots & 0 & -g\sqrt{2s}\\
0 & \cdots & 0 & 0
\end{array}\right)}}\\
 &  & \ddots
\end{array}\right).
\end{align}
Therefore, the matrix representation of the complete Hamiltonian $\hat{H}$
is given by 
\begin{equation}
H=\stackrel[n=0]{n_{\max}}{\oplus}\left(E_{n}I_{2s_{n}+1}+\mathbb{J}_{2s_{n}+1}\right).
\end{equation}
Without loss of generality, we can order $E_{n}$ such that $E_{0}$
is the lowest energy and $E_{n\neq0}-E_{0}\neq0$. The rank of $H-E_{0}I_{2^{L}}$
can be shown to be 
\begin{equation}
\text{rank}\left(H-E_{0}I_{2^{L}}\right)=2^{L}-1.
\end{equation}
Given that $\mathbb{J}_{2s_{n}+1}$ is a nilpotent matrix satisfying
$(\mathbb{J}_{2s_{n}+1})^{2s_{n}+1}=\mathbf{0}_{2s_{n}+1}$, we can
further deduce that 
\begin{equation}
\text{rank}\left((H-E_{0}I_{2^{L}})^{p}\right)=\begin{cases}
2s_{n}+1-p, & p\leq2s_{0}+1\\
0 & p>2s_{0}+1
\end{cases}.
\end{equation}

Provided the invertible matrix $S$ transforms $H$ to its Jordan
normal form, denoted as $S^{-1}HS=J_{H}$, we find

\begin{equation}
S^{-1}(H-E_{0}I_{2^{L}})^{p}S=(J_{H}-\stackrel[n=0]{n_{\max}}{\oplus}E_{0}I_{2s_{n}+1})^{p},
\end{equation}
which yields

\begin{align}
 & \mathrm{rank}\left((J_{H}-\stackrel[n=0]{n_{\max}}{\oplus}E_{0}I_{2s_{n}+1})^{p}\right)\nonumber \\
= & \mathrm{rank}\left((H-E_{0}I_{2^{L}})^{p}\right)\nonumber \\
= & \begin{cases}
2s_{n}+1-p, & p\leq2s_{0}+1\\
0 & p>2s_{0}+1
\end{cases}.
\end{align}
The unique ansatz of the Jordan normal matrix satisfying the above
relation reads

\begin{equation}
J_{H}=J_{H}^{\prime}\oplus J_{2s_{0}+1}(E_{0}),
\end{equation}
where $J^{\prime}$ is the sub-block of the remainder. This result
implies that the effective Hamiltonian has an EP of order $2s_{0}+1$
as given by the Jordan block $J_{2s_{0}+1}(E_{0})$. In the case of
a finite size $L$ spin-$1/2$ system, $s_{0}$ corresponding to the
ground state of the spin-$\frac{1}{2}$ Heisenberg XXX chain depends
on the system size, i.e., $s_{0}=L/2$, leading to an EP of order
$L+1$.

For the Heisenberg spin chain under the periodic boundary condition,
as shown in the main text, an additional index is needed to label
the eigenvalue of the translation operator. However, in the ground
state $|n=0\rangle$, the translation operator assumes only one eigenvalue,
since $|n=0\rangle$ inherently satisfies translation invariance.
As a result, the periodic boundary condition does not change the EP
of the highest order found in the open boundary condition.

\subsubsection{Arbitrary Local-Site Defective Term }

Next, we discuss the case where the symmetry-breaking term and the
measurement are applied simultaneously to those local sites in $\mathcal{M}$.
Here, the defective term, $-g\sum_{j\in\mathcal{M}}\hat{S}_{j}^{-}$,
cannot be expressed using total spin operators. It can, however, be
expressed as the partial spin operator $-g\hat{S}_{1}^{-}$, with
$\hat{S}_{1}^{x(y)(z)}\equiv\sum_{j\in\mathcal{M}}\hat{S}_{j}^{x(y)(z)}$.
Furthermore, we introduce the partial spin operator for the remaining
part, $\hat{S}_{2}^{x(y)(z)}\equiv\sum_{j\notin\mathcal{M}}\hat{S}_{j}^{x(y)(z)}$.
Thus, the defective term, $-g\hat{S}_{1}^{-}$, can be fully expressed
as

\begin{equation}
-g\hat{S}_{1}^{-}\otimes\hat{I}_{2}.
\end{equation}
In the decoupled representation with respect to $\{|s_{1},m_{s_{1}};s_{2},m_{s_{2}}\rangle\}$,
the defective term's matrix elements are 

\begin{align}
 & \langle s_{1}^{\prime},m_{s_{1}^{\prime}}^{\prime};s_{2}^{\prime},m_{s_{2}^{\prime}}^{\prime}|\hat{S}_{1}^{-}\otimes\hat{I}_{2}|s_{1},m_{s_{1}};s_{2},m_{s_{2}}\rangle\nonumber \\
\propto & \delta_{s_{1}^{\prime},s_{1}}\delta_{m_{s_{1}^{\prime}}^{\prime},m_{s_{1}}-1}\delta_{s_{2}^{\prime},s_{2}}\delta_{m_{s_{2}^{\prime}}^{\prime},m_{s_{2}}}.
\end{align}
In the coupled basis, $\{|s,m_{s}\rangle\}$, the matrix elements
are given by \begin{widetext}
\begin{align}
 & \langle s^{\prime},m_{s^{\prime}}^{\prime}|\hat{S}_{1}^{-}\otimes\hat{I}_{2}|s,m_{s}\rangle\nonumber \\
= & \sum_{\begin{array}{c}
s_{1}^{\prime},m_{s_{1}^{\prime}}^{\prime},s_{2}^{\prime},m_{s_{2}^{\prime}}^{\prime}\\
s_{1},m_{s_{1}},s_{2},m_{s_{2}}
\end{array}}\langle s^{\prime},m_{s^{\prime}}^{\prime}|s_{1}^{\prime},m_{s_{1}^{\prime}}^{\prime};s_{2}^{\prime},m_{s_{2}^{\prime}}^{\prime}\rangle\langle s_{1}^{\prime},m_{s_{1}^{\prime}}^{\prime};s_{2}^{\prime},m_{s_{2}^{\prime}}^{\prime}|\hat{S}_{1}^{-}\otimes\hat{I}_{2}|s_{1},m_{s_{1}};s_{2},m_{s_{2}}\rangle\langle s_{1},m_{s_{1}};s_{2},m_{s_{2}}|s,m_{s}\rangle\nonumber \\
= & \sum_{s_{1},m_{s_{1}},s_{2},m_{s_{2}}}\sqrt{(s_{1}+m_{s_{1}})(s_{1}-m_{s_{1}}+1)}\langle s^{\prime},m_{s^{\prime}}^{\prime}|s_{1},m_{s_{1}}-1;s_{2},m_{s_{2}}\rangle\langle s_{1},m_{s_{1}};s_{2},m_{s_{2}}|s,m_{s}\rangle.
\end{align}
\end{widetext}From the basic properties of Clebsch-Gordan coefficients,
we know that 
\begin{equation}
m_{s}=m_{s_{1}}+m_{s_{2}},\,\,m_{s^{\prime}}^{\prime}=m_{s_{1}}-1+m_{s_{2}}
\end{equation}
must be satisfied for each non-zero term in the above summation. This
implies that $m_{s^{\prime}}^{\prime}=m_{s}-1$ must be satisfied
for non-zero matrix element of $-g\hat{S}_{1}^{-}\otimes\hat{I}_{2}$
in the coupled basis, yielding
\begin{equation}
\langle s^{\prime},m_{s^{\prime}}^{\prime}|(-g\hat{S}_{1}^{-}\otimes\hat{I}_{2})|s,m_{s}\rangle\propto\delta_{m_{s^{\prime}}^{\prime},m_{s}-1}.
\end{equation}
Let's denote the matrix representation of $-g\hat{S}_{1}^{-}\otimes\hat{I}_{2}$
with respect to the basis $\{|n,s_{n},m_{s}\rangle\}$ as $\mathbb{M}^{(1)}$.
The matrix elements in row $(n^{\prime},s_{n^{\prime}}^{\prime},m_{s^{\prime}}^{\prime})$,
column $(n,s_{n},m_{s})$ are given by

\begin{equation}
\mathbb{M}_{(n^{\prime},s_{n^{\prime}}^{\prime},m_{s^{\prime}}^{\prime}),(n,s_{n},m_{s})}^{(1)}=C_{(n^{\prime},s_{n^{\prime}}^{\prime},m_{s^{\prime}}^{\prime}),(n,s_{n},m_{s})}^{(1)}\delta_{m_{s^{\prime}}^{\prime},m_{s}-1}\label{eq:matrix_element_M1}
\end{equation}
where $C_{(n^{\prime},s_{n^{\prime}}^{\prime},m_{s^{\prime}}^{\prime}),(n,s_{n},m_{s})}^{(1)}$
is a coefficient. Therefore the complete effective Hamiltonian has
the matrix representation

\begin{equation}
H\equiv\stackrel[n=0]{n_{\max}}{\oplus}E_{n}I_{2s_{n}+1}+\mathbb{M}^{(1)}.
\end{equation}

From this point, we aim to demonstrate that the Jordan normal matrix
of $H$ is given by $J_{2s_{0}+1}(E_{0})\oplus J^{\prime}$, where
$J^{\prime}$ denotes the sub-block corresponding to the remainder.
We reorder the index such that $s_{0}=\max(\{s_{n}\})$ for convenience
in our argument. We begin by verifying that $E_{0}$ is indeed an
eigenvalue of $\hat{H}$ with respect to the eigenstate $|n=0,s_{n}=s_{0},m_{s}=-s_{0}\rangle$,
\begin{align}
 & \hat{H}|0,s_{0},-s_{0}\rangle\nonumber \\
= & \sum_{n^{\prime},s_{n^{\prime}}^{\prime},m_{s^{\prime}}^{\prime}}|n^{\prime},s_{n^{\prime}}^{\prime},m_{s^{\prime}}^{\prime}\rangle H_{(n^{\prime},s_{n^{\prime}}^{\prime},m_{s^{\prime}}^{\prime}),(0,s_{0},-s_{0})}\nonumber \\
= & \sum_{n^{\prime},s_{n^{\prime}}^{\prime},m_{s^{\prime}}^{\prime}}|n^{\prime},s_{n^{\prime}}^{\prime},m_{s^{\prime}}^{\prime}\rangle E_{n^{\prime}}\delta_{n^{\prime},0}\delta_{s_{n^{\prime}}^{\prime},s_{0}}\delta_{m_{s^{\prime}}^{\prime},-s_{0}}\nonumber \\
 & +\sum_{n^{\prime},s_{n^{\prime}}^{\prime},m_{s^{\prime}}^{\prime}}|n^{\prime},s_{n^{\prime}}^{\prime},m_{s^{\prime}}^{\prime}\rangle C_{(n^{\prime},s_{n^{\prime}}^{\prime},m_{s^{\prime}}^{\prime}),(0,s_{0},-s_{0})}\delta_{m_{s^{\prime}}^{\prime},-s_{0}-1}\nonumber \\
= & E_{0}|0,s_{0},-s_{0}\rangle.\label{eq:Heisenberg_EP_eigenvector}
\end{align}
The last step employs $s_{0}=\max(\{s_{n}\})$, implying that $\delta_{m_{s^{\prime}}^{\prime},-s_{0}-1}$
does not contribute to the summation. Given that each matrix can be
similar to a matrix in Jordan normal form and that similarity transformation
maintains the rank invariant, we can directly compute the rank of
$\tilde{H}=H-\stackrel[n=0]{n_{\max}}{\oplus}E_{0}I_{2s_{n}+1}$ to
various powers to derive the characteristics of the Jordan sub-block
associated with $|n=0\rangle$. The matrix of $\tilde{H}$ can be
expressed as 
\begin{equation}
\left(\begin{array}{cc}
\ddots & \vdots\\
\cdots & \overset{(2s_{0}+1)\times(2s_{0}+1)}{\overbrace{\left(\begin{array}{ccccc}
0 & * & 0 & \cdots & 0\\
\vdots & \ddots & * & \iddots & \vdots\\
0 & \cdots & 0 & * & 0\\
\vdots & \iddots & \vdots & \ddots & *\\
0 & \cdots & 0 & \cdots & 0
\end{array}\right)}}
\end{array}\right),
\end{equation}
where $*$ represents the nonzero off-diagonal elements as defined
by (\ref{eq:matrix_element_M1}). Due to the presence of cross block
terms, deciphering the rank of the matrix $\tilde{H}$ directly poses
a challenge. Consequently, we employ some operations to reveal the
rank.

Note that if $B$ is an invertible square matrix, the rank of square
matrix $A$ satisfies 
\begin{equation}
\text{rank}\left(A\right)=\text{rank}\left(AB\right)=\text{rank}\left(BA\right).
\end{equation}
We consider those $B$ of the elementary matrix such that $BA$ represents
the elementary row operations, and $AB$ stands for the elementary
column operations. In the case of $\tilde{H}$, all elements of row
$(0,s_{0},s_{0})$ are zeros since no column with index $m_{s}=s_{0}+1$
exists, implying that the rank of $\tilde{H}$ can at most be $d-1$
where $d=\sum_{n=0}^{n_{max}}\left(2s_{n}+1\right)=2^{L}$. To verify
this, we apply elementary column operations to eliminate some elements
of the remaining rows of $|n=0\rangle$. For example, we use the nonzero
$*$ in column $(0,s_{0},m_{s}=s_{0},s_{0}-1,\cdots,-s_{0}+1)$ to
cancel all elements in row $(0,s_{0},m_{s}-1)$. After this operation,
the matrix $\tilde{H}$ transforms into 
\begin{equation}
\left(\begin{array}{cc}
\ddots & \vdots\\
\left(\begin{array}{ccccc}
0 & \cdots & 0 & \cdots & 0\\
\vdots & \vdots & \vdots & \vdots & \vdots\\
0 & \cdots & 0 & \cdots & 0\\
\vdots & \vdots & \vdots & \vdots & \vdots\\
0 & \cdots & 0 & \cdots & 0
\end{array}\right) & \overset{(2s_{0}+1)\times(2s_{0}+1)}{\overbrace{\left(\begin{array}{ccccc}
0 & * & 0 & \cdots & 0\\
\vdots & \ddots & * & \iddots & \vdots\\
0 & \cdots & 0 & * & 0\\
\vdots & \iddots & \vdots & \ddots & *\\
0 & \cdots & 0 & \cdots & 0
\end{array}\right)}}
\end{array}\right).
\end{equation}
Subsequently, we can employ elementary row operations to eliminate
all the elements in the upper-right block using the nonzero $*$'s,
leading to
\begin{equation}
\left(\begin{array}{cc}
\ddots & \left(\begin{array}{ccccc}
0 & \cdots & 0 & \cdots & 0\\
\vdots & \vdots & \vdots & \vdots & \vdots\\
0 & \cdots & 0 & \cdots & 0\\
\vdots & \vdots & \vdots & \vdots & \vdots\\
0 & \cdots & 0 & \cdots & 0
\end{array}\right)\\
\left(\begin{array}{ccccc}
0 & \cdots & 0 & \cdots & 0\\
\vdots & \vdots & \vdots & \vdots & \vdots\\
0 & \cdots & 0 & \cdots & 0\\
\vdots & \vdots & \vdots & \vdots & \vdots\\
0 & \cdots & 0 & \cdots & 0
\end{array}\right) & \overset{(2s_{0}+1)\times(2s_{0}+1)}{\overbrace{\left(\begin{array}{ccccc}
0 & * & 0 & \cdots & 0\\
\vdots & \ddots & * & \iddots & \vdots\\
0 & \cdots & 0 & * & 0\\
\vdots & \iddots & \vdots & \ddots & *\\
0 & \cdots & 0 & \cdots & 0
\end{array}\right)}}
\end{array}\right),
\end{equation}
where the column $(0,s_{0},-s_{0})$ is populated with zeros since
no row with $m_{s^{\prime}}^{\prime}=-s_{0}-1$ exists. Importantly,
the diagonal terms in the upper-left $(d-2s_{0}-1)\times(d-2s_{0}-1)$
block remain unaffected during the preceding elementary column operations.
This can be seen as the column $(n,s_{n},m_{s})$ possesses a nonzero
element in row $(0,s_{0},m_{s}-1)$, and the $*$ utilized to cancel
it is in column $(0,s_{0},m_{s})$, with $\mathbb{M}_{(0,s_{0},m_{s}),(0,s_{0},m_{s})}^{(1)}=0$.
Therefore, using these diagonal terms in the upper-left block to cancel
all the remaining elements in the same block by either elementary
row or column operations, the matrix $\tilde{H}$ transforms into
\begin{equation}
\overset{(d-2s_{0}-1)\times(d-2s_{0}-1)}{\overbrace{\left(\begin{array}{ccc}
E_{n_{max}}-E_{0} & \cdots & 0\\
\vdots & \ddots & \vdots\\
0 & \cdots & E_{1}-E_{0}
\end{array}\right)}}\oplus\overset{(2s_{0}+1)\times(2s_{0}+1)}{\overbrace{\left(\begin{array}{ccccc}
0 & * & 0 & \cdots & 0\\
\vdots & \ddots & * & \iddots & \vdots\\
0 & \cdots & 0 & * & 0\\
\vdots & \iddots & \vdots & \ddots & *\\
0 & \cdots & 0 & \cdots & 0
\end{array}\right)}}.
\end{equation}
We can now determine the rank to be
\begin{align}
\mathrm{rank}\left(\tilde{H}\right) & =d-2s_{0}-1+2s_{0}\nonumber \\
 & =d-1.\label{eq:rank_p=00003D1}
\end{align}

To investigate the rank for $p>1$, we define the linear mapping $f:\mathbb{C}^{d}\rightarrow\mathbb{C}^{d}$
as 
\begin{align}
f(\mathbf{x}) & =\tilde{H}\mathbf{x},
\end{align}
where 
\begin{equation}
\mathbf{x}=\left(\begin{array}{c}
x_{(n_{\max},s_{n_{\max}},-s_{n_{\max}})}\\
x_{(n_{\max},s_{n_{\max}},-s_{n_{\max}}+1)}\\
\vdots\\
x_{(0,s_{0},s_{0})}
\end{array}\right)\in\mathbb{C}^{d}
\end{equation}
is a $d\times1$ column vector, with each row element being an independent
complex scalar. The rank of $\tilde{H}$ is then given by 
\begin{align}
\mathrm{rank}\left(\tilde{H}\right) & =\dim_{\mathbb{C}}(f(\mathbb{C}^{d})),
\end{align}
where the image of $f$ is defined by 
\begin{align}
f(\mathbb{C}^{d}) & =\{f(\mathbf{x})|\mathbf{x}\in\mathbb{C}^{d}\}.
\end{align}
For ease of argument, we express column vectors in $\tilde{H}$ as
\[
\mathbf{v}_{(n,s_{n},m_{s})}=\left(\begin{array}{c}
\tilde{H}_{(n_{\max},s_{n_{\max}},-s_{n_{\max}}),(n,s_{n},m_{s})}\\
\tilde{H}_{(n_{\max},s_{n_{\max}},-s_{n_{\max}}+1),(n,s_{n},m_{s})}\\
\vdots\\
\tilde{H}_{(0,s_{0},s_{0}),(n,s_{n},m_{s})}
\end{array}\right)
\]
such that 
\begin{align}
f(\mathbb{C}^{d}) & =\text{span}\left\{ \mathbf{v}_{(n_{\max},s_{n_{\max}},-s_{n_{\max}})},\dots,\mathbf{v}_{(0,s_{0},s_{0})}\right\} .
\end{align}

We begin by revisiting the case of $p=1$ to understand what the result
in (\ref{eq:rank_p=00003D1}) implies in the context of the linear
mapping $f$. Noting that $\tilde{H}_{(n^{\prime},s_{n^{\prime}}^{\prime},m_{s^{\prime}}^{\prime}),(0,s_{0},-s_{0})}=\tilde{H}_{(0,s_{0},s_{0}),(n,s_{n},m_{s})}=0$
(due to $\delta_{m_{s^{\prime}}^{\prime},m_{s}-1}$ in (\ref{eq:matrix_element_M1})),
we find that $f(\mathbf{x})$ neither includes $x_{(0,s_{0},-s_{0})}$
nor contains a nonzero value in row $(0,s_{0},s_{0})$, i.e., 
\begin{align}
f(\mathbb{C}^{d})=\text{span} & \left\{ \mathbf{v}_{(n_{\max},s_{n_{\max}},-s_{n_{\max}})},\dots,\mathbf{v}_{(1,s_{1},s_{1})},\right.\nonumber \\
 & \left.\mathbf{v}_{(0,s_{0},-s_{0}+1)},\dots,\mathbf{v}_{(0,s_{0},s_{0})}\right\} ,
\end{align}
\begin{align}
 & f(\mathbf{x})\nonumber \\
= & \left(\begin{array}{c}
\underset{(n,s_{n},m_{s})}{\sum}\tilde{H}_{(n_{\max},s_{n_{\max}},-s_{n_{\max}}),(n,s_{n},m_{s})}x_{(n,s_{n},m_{s})}\\
\vdots\\
\underset{(n,s_{n},m_{s})}{\sum}\tilde{H}_{(0,s_{0},s_{0}-1),(n,s_{n},m_{s})}x_{(n,s_{n},m_{s})}\\
0
\end{array}\right).
\end{align}
This equation implies that $f(\mathbb{C}^{d})$ is spanned by $d-1$
vectors, resulting in $\dim_{\mathbb{C}}(f(\mathbb{C}^{d}))\leq d-1$.
From the result in (\ref{eq:rank_p=00003D1}), i.e., 
\begin{equation}
\dim_{\mathbb{C}}(f(\mathbb{C}^{d}))=d-1,
\end{equation}
we infer that $\mathbf{v}_{(n_{\max},s_{n_{\max}},-s_{n_{\max}})}$,
$\mathbf{v}_{(n_{\max},s_{n_{\max}},-s_{n_{\max}}+1)}$, $\dots$,
$\mathbf{v}_{(0,s_{0},s_{0}-1)}$ form a basis for $f(\mathbb{C}^{d})$,
implying they are linearly independent. This also suggests that we
require $d-1$ independent complex scalars as the coordinates of the
vector $f(\mathbf{x})$ within a basis. Consequently, we introduce
the following compact notation, 
\begin{align}
\mathbf{x}^{(1)} & \equiv f(\mathbf{x})\nonumber \\
 & =\left(\begin{array}{c}
x_{(n_{\max},s_{n_{\max}},-s_{n_{\max}})}^{(1)}\\
\vdots\\
x_{(0,s_{0},s_{0}-1)}^{(1)}\\
0
\end{array}\right),
\end{align}
where $x_{(n,s_{n},m_{s})\neq(0,s_{0},s_{0})}^{(1)}\in\mathbb{C}$.
Therefore, we find that the $d-1$ complex scalars in $\mathbf{x}^{(1)}$
must be independent and can be regarded as the starting point for
the subsequent steps. 

Before generalizing to the case of an arbitrary $p$, let us consider
$p=2$ and examine how the structure of $f(\mathbf{x})$ is extended
to $f(\mathbf{x}^{(1)})$. Observing that $f(f(\mathbb{C}^{d}))\subseteq f(\mathbb{C}^{d})$
($f(\mathbb{C}^{d})\subseteq\mathbb{C}^{d}$), and considering that
there are only $d-1$ independent complex scalars in $\mathbf{x}^{(1)}$,
we conclude that $f(\mathbf{x}^{(1)})$ does not contain $x_{(0,s_{0},-s_{0})}^{(1)}$
and $x_{(0,s_{0},s_{0})}^{(1)}$, i.e., 
\begin{align}
f(f(\mathbb{C}^{d}))=\text{span} & \left\{ \mathbf{v}_{(n_{\max},s_{n_{\max}},-s_{n_{\max}})},\dots,\mathbf{v}_{(1,s_{1},s_{1})},\right.\nonumber \\
 & \left.\mathbf{v}_{(0,s_{0},-s_{0}+1)},\dots,\mathbf{v}_{(0,s_{0},s_{0}-1)}\right\} ,
\end{align}
\begin{align}
 & f(\mathbf{x}^{(1)})\nonumber \\
= & \left(\begin{array}{c}
\underset{(n,s_{n},m_{s})}{\sum}\tilde{H}_{(n_{\max},s_{n_{\max}},-s_{n_{\max}}),(n,s_{n},m_{s})}x_{(n,s_{n},m_{s})}^{(1)}\\
\vdots\\
\underset{(n,s_{n},m_{s})}{\sum}\tilde{H}_{(0,s_{0},s_{0}-1),(n,s_{n},m_{s})}x_{(n,s_{n},m_{s})}^{(1)}\\
0
\end{array}\right).
\end{align}
Given that the above $d-2$ vectors are combinations of the basis
vectors of $f(\mathbb{C}^{d})$, they must be linearly independent
and hence form the basis of $f(f(\mathbb{C}^{d}))$, which implies
\begin{equation}
\dim_{\mathbb{C}}(f(f(\mathbb{C}^{d})))=d-2.
\end{equation}
This indicates that only $d-2$ complex scalars are needed to characterize
the vector $f(\mathbf{x}^{(1)})$. It is important to note that the
sole difference between $\mathbf{x}^{(1)}$ and $\mathbf{x}$ is the
element in row $(0,s_{0},s_{0})$. Moreover, only row $(n^{\prime},s_{n^{\prime}}^{\prime},m_{s^{\prime}}^{\prime}=s_{0}-1)$
is nonzero in column $(0,s_{0},s_{0})$ of $\tilde{H}$. These observations
suggest that only the rows $(n^{\prime},s_{n^{\prime}}^{\prime},m_{s^{\prime}}^{\prime}=s_{0}-1)$
of $f(\mathbf{x}^{(1)})$ differ from those of $f(\mathbf{x})$. From
the fact that the $(d-2s_{0}-1)\times(d-2s_{0}-1)$ sub-block of $\tilde{H}$
has full rank (owing to the presence of the nonzero diagonal terms),
we infer that the complex scalars in rows $(n^{\prime}\neq0,s_{n^{\prime}}^{\prime},m_{s^{\prime}}^{\prime})$
of $f(\mathbf{x}^{(1)})$ are independent. As a result, the row $(0,s_{0},s_{0}-1)$
of $f(\mathbf{x}^{(1)})$ must depend on other rows. To succinctly
express this, we introduce the following compact notation, 
\begin{align}
 & \mathbf{x}^{(2)}\equiv f(\mathbf{x}^{(1)})\nonumber \\
= & \left(\begin{array}{c}
x_{(n_{\max},s_{n_{\max}},-s_{n_{\max}})}^{(2)}\\
\vdots\\
x_{(0,s_{0},s_{0}-2)}^{(2)}\\
\underset{\begin{array}{c}
\{(n^{\prime}\neq0,s_{n^{\prime}},m_{s^{\prime}}^{\prime})\}\\
\cup\{(0,s_{0},m_{s^{\prime}}^{\prime}<s_{0}-1)\}
\end{array}}{\sum}c_{(n^{\prime},s_{n^{\prime}}^{\prime},m_{s^{\prime}}^{\prime})}^{(2),2}x_{(n^{\prime},s_{n^{\prime}}^{\prime},m_{s^{\prime}}^{\prime})}^{(2)}\\
0
\end{array}\right).
\end{align}
Here, the $d-2$ complex scalars in $\mathbf{x}^{(2)}$ are independent
and can be regarded as the starting point for the next step.

For $2<p<2s_{0}+2$, the vector in $\overset{p-1}{\overbrace{f(f(\dots f}}(\mathbb{C}^{d})\dots))$
reads 
\begin{align}
 & \mathbf{x}^{(p-1)}\nonumber \\
= & \left(\begin{array}{c}
x_{(n_{\max},s_{n_{\max}},-s_{n_{\max}})}^{(p-1)}\\
\vdots\\
x_{(0,s_{0},s_{0}-(p-1))}^{(p-1)}\\
\underset{\begin{array}{c}
\{(n^{\prime}\neq0,s_{n^{\prime}},m_{s^{\prime}}^{\prime})\}\\
\cup\{(0,s_{0},m_{s^{\prime}}^{\prime}<s_{0}-p+1)\}
\end{array}}{\sum}c_{(n^{\prime},s_{n^{\prime}}^{\prime},m_{s^{\prime}}^{\prime})}^{(p-1),p-1}x_{(n^{\prime},s_{n^{\prime}}^{\prime},m_{s^{\prime}}^{\prime})}^{(p-1)}\\
\vdots\\
\underset{\begin{array}{c}
\{(n^{\prime}\neq0,s_{n^{\prime}},m_{s^{\prime}}^{\prime})\}\\
\cup\{(0,s_{0},m_{s^{\prime}}^{\prime}<s_{0}-p+1)\}
\end{array}}{\sum}c_{(n^{\prime},s_{n^{\prime}}^{\prime},m_{s^{\prime}}^{\prime})}^{(2),p-1}x_{(n^{\prime},s_{n^{\prime}}^{\prime},m_{s^{\prime}}^{\prime})}^{(p-1)}\\
0
\end{array}\right).\label{eq:vect_p-1}
\end{align}
Observing that $\overset{p}{\overbrace{f(f(\dots f}}(\mathbb{C}^{d})\dots))\subseteq\overset{p-1}{\overbrace{f(f(\dots f}}(\mathbb{C}^{d})\dots))$,
and considering that there are only $d-p+1$ independent complex scalars
in $\mathbf{x}^{(p-1)}$, we conclude that $f(\mathbf{x}^{(p-1)})$
does not contain $x_{(0,s_{0},-s_{0})}^{(p-1)}$ and $x_{(0,s_{0},m_{s^{\prime}}^{\prime}=s_{0}-(p-2),\dots,s_{0})}^{(p-1)}$,
i.e., 
\begin{align}
 & \overset{p}{\overbrace{f(f(\dots f}}(\mathbb{C}^{d})\dots))\nonumber \\
= & \text{span}\left\{ (\mathbf{v}_{(n_{\max},s_{n_{\max}},-s_{n_{\max}})}+*),\dots,(\mathbf{v}_{(1,s_{1},s_{1})}+*),\right.\nonumber \\
 & \left.(\mathbf{v}_{(0,s_{0},-s_{0}+1)}+*),\dots,(\mathbf{v}_{(0,s_{0},s_{0}-(p-1))}+*)\right\} .
\end{align}
In the above expression, different $*$'s denote different combinations
of $\mathbf{v}_{(0,s_{0},m_{s^{\prime}}^{\prime}\geq s_{0}-p+1)}$
with coefficients provided in (\ref{eq:vect_p-1}). The above $d-p$
linearly independent vectors yield 
\begin{equation}
\dim_{\mathbb{C}}(\overset{p}{\overbrace{f(f(\dots f}}(\mathbb{C}^{d})\dots)))=d-p.
\end{equation}
This implies that only $d-p$ complex scalars are needed to characterize
the vector $f(\mathbf{x}^{(p-1)})$. It's worth noting that the only
difference between $\mathbf{x}^{(p-1)}$ and $\mathbf{x}^{(p-2)}$
is the element in row $(0,s_{0},s_{0}-(p-2))$. Additionally, only
row $(n^{\prime},s_{n^{\prime}}^{\prime},m_{s^{\prime}}^{\prime}=s_{0}-(p-1))$
is nonzero in column $(0,s_{0},s_{0}-(p-2))$ of $\tilde{H}$. These
observations indicate that only the rows $(n^{\prime},s_{n^{\prime}}^{\prime},m_{s^{\prime}}^{\prime}=s_{0}-(p-1))$
of $f(\mathbf{x}^{(p-1)})$ differ from those of $f(\mathbf{x}^{(p-2)})$.
Given that the $(d-2s_{0}-1)\times(d-2s_{0}-1)$ sub-block of $\tilde{H}$
has full rank, we infer that the complex scalars in rows $(n^{\prime}\neq0,s_{n^{\prime}}^{\prime},m_{s^{\prime}}^{\prime})$
of $f(\mathbf{x}^{(p-1)})$ are independent. Consequently, we deduce
that the row $(0,s_{0},s_{0}-(p-1))$ of $f(\mathbf{x}^{(p-1)})$
must depend on other rows. Therefore, we introduce 
\begin{align}
 & \mathbf{x}^{(p)}\equiv f(\mathbf{x}^{(p-1)})\nonumber \\
= & \left(\begin{array}{c}
x_{(n_{\max},s_{n_{\max}},-s_{n_{\max}})}^{(p)}\\
\vdots\\
x_{(0,s_{0},s_{0}-p)}^{(p)}\\
\underset{\begin{array}{c}
\{(n^{\prime}\neq0,s_{n^{\prime}},m_{s^{\prime}}^{\prime})\}\\
\cup\{(0,s_{0},m_{s^{\prime}}^{\prime}<s_{0}-p)\}
\end{array}}{\sum}c_{(n^{\prime},s_{n^{\prime}}^{\prime},m_{s^{\prime}}^{\prime})}^{(p),p}x_{(n^{\prime},s_{n^{\prime}}^{\prime},m_{s^{\prime}}^{\prime})}^{(p)}\\
\vdots\\
\underset{\begin{array}{c}
\{(n^{\prime}\neq0,s_{n^{\prime}},m_{s^{\prime}}^{\prime})\}\\
\cup\{(0,s_{0},m_{s^{\prime}}^{\prime}<s_{0}-p)\}
\end{array}}{\sum}c_{(n^{\prime},s_{n^{\prime}}^{\prime},m_{s^{\prime}}^{\prime})}^{(2),p}x_{(n^{\prime},s_{n^{\prime}}^{\prime},m_{s^{\prime}}^{\prime})}^{(p)}\\
0
\end{array}\right),
\end{align}
where the $d-p$ complex scalars in $\mathbf{x}^{(p)}$ are independent
and can be regarded as the starting point for the next step. 

For $p\geq2s_{0}+2$, we can expect $\overset{2s_{0}+2}{\overbrace{f(f(\dots f}}(\mathbb{C}^{d})\dots))=\overset{2s_{0}+1}{\overbrace{f(f(\dots f}}(\mathbb{C}^{d})\dots))$
since $\tilde{H}_{(n^{\prime},s_{n^{\prime}}^{\prime},m_{s^{\prime}}^{\prime}),(0,s_{0},-s_{0})}=0$.
This leads to
\begin{align}
\dim_{\mathbb{C}}(\overset{2s_{0}+2}{\overbrace{f(f(\dots f}}(\mathbb{C}^{d})\dots))) & =\dim_{\mathbb{C}}(\overset{2s_{0}+1}{\overbrace{f(f(\dots f}}(\mathbb{C}^{d})\dots)))\nonumber \\
 & =d-2s_{0}-1.
\end{align}

Ultimately, we derive 
\begin{align}
\mathrm{rank}\left(\tilde{H}^{p}\right) & =\dim_{\mathbb{C}}(\overset{p}{\overbrace{f(f(\dots f}}(\mathbb{C}^{d})\dots)))\nonumber \\
 & =\begin{cases}
d-p, & p\leq2s_{0}+1\\
d-2s_{0}-1 & p>2s_{0}+1
\end{cases}.
\end{align}
Provided that the invertible matrix $S$ transforms $H$ to its Jordan
normal form, i.e., $S^{-1}HS=J_{H}$, we find 
\begin{align}
 & \mathrm{rank}\left((J_{H}-\stackrel[n=0]{n_{\max}}{\oplus}E_{0}I_{2s_{n}+1})^{p}\right)\nonumber \\
= & \mathrm{rank}\left(\tilde{H}^{p}\right)\nonumber \\
= & \begin{cases}
d-p, & p\leq2s_{0}+1\\
d-2s_{0}-1 & p>2s_{0}+1
\end{cases}.
\end{align}
The only plausible Jordan normal matrix satisfying the above relation
is 
\begin{equation}
J_{H}=J_{H}^{\prime}\oplus J_{2s_{0}+1}(E_{0}),
\end{equation}
where $J_{H}^{\prime}$ represents the remaining sub-block. We now
find that the effective Hamiltonian possesses an EP of order $2s_{0}+1$,
as implied by the Jordan block $J_{2s_{0}+1}(E_{0})$. For a finite-sized
spin-$1/2$ system, where $L$ represents the size, the $s_{0}$ associated
with the ground state of the Heisenberg XXX chain depends on the system
size, $s_{0}=L/2$. This results in an EP of order $L+1$.

In addition, the computation detailed above can be adjusted to accommodate
the periodic boundary condition. As previously noted in the discussion
of the global measurement case, this adjustment involves appending
one more index to label the eigenvalue of the translation operator.
The crucial structure and the final result, i.e., maximum power with
respect to $t$, remain unaltered, as the translation operator assumes
only one eigenvalue in the maximally polarized state $|n=0\rangle$.

\subsection{Ising Chain}

In order to determine the existence and order of the EPs in $\hat{H}_{\mathrm{TI}}^{\mathrm{eff}}$,
we first analyze the eigensystem of the Ising model. For the Ising
model with periodic boundary conditions, the Hamiltonian is 
\begin{equation}
\hat{H}_{\text{I}}=-J\sum_{j=1}^{L}\hat{\sigma}_{j}^{z}\hat{\sigma}_{j+1}^{z},\,(\hat{\sigma}_{L+1}^{z}=\hat{\sigma}_{1}^{z})
\end{equation}
The set of all eigenstates is given by $\{|\sigma_{1}\sigma_{2}\cdots\sigma_{L}\rangle\}$,
where $\sigma\in\{\uparrow,\downarrow\}$, and $|\uparrow_{j}\rangle$
($|\downarrow_{j}\rangle$) is the eigenstate of $\hat{\sigma}_{j}^{z}$
with an eigenenergy of $+1$ ($-1$). We treat the all-spin-down polarized
state, $|\downarrow_{1}\downarrow_{2}\cdots\downarrow_{L}\rangle$,
as the vacuum, and for convenience, write the one-magnon states as
$|\mu\rangle\equiv|\downarrow_{1}\downarrow_{2}\cdots\uparrow_{\mu}\cdots\downarrow_{L}\rangle$.
Similarly, the $n$-magnon states ($L-2\geq n\geq2$) are represented
as 
\begin{equation}
|\mu_{1},\cdots,\mu_{n}\rangle\equiv|\downarrow_{1}\cdots\downarrow_{\mu_{1}-1}\uparrow_{\mu_{1}}\cdots\uparrow_{\mu_{n}}\downarrow_{\mu_{n}+1}\cdots\downarrow_{L}\rangle.
\end{equation}
Given that the operator for the total number of domain walls can be
expressed by the Hamiltonian as $\hat{D}\equiv\frac{L}{2}+\frac{\hat{H}_{\text{I}}}{2J}$,
we can denote the energy eigenstate by the total number of domain
walls $D\equiv\langle\hat{D}\rangle$. In particular, $|D;\mu_{1},\mu_{2},\cdots,\mu_{n}\rangle$
denotes the $n$-magnon state that has $D$ domain walls and the same
eigenenergy $E_{D}=J(2D-L)$. For example, all the states $|D=2;\mu,\mu+1,\cdots,\mu+n-1\rangle$
($L-2\geq n\geq1$, $\mu\in\{1,2,\cdots,L\}$) share the same energy
$J(4-L)$, i.e., 
\begin{align}
 & \hat{H}_{\text{I}}|2;\mu,\mu+1,\cdots,\mu+n-1\rangle\nonumber \\
= & J(4-L)|2;\mu,\mu+1,\cdots,\mu+n-1\rangle.
\end{align}
In the following, we will examine how the local-site defective components,
$\hat{\sigma}_{j}^{x}-i\hat{\sigma}_{j}^{y}$, affect the eigensystem
of $\hat{H}_{\text{I}}$.

\subsubsection{Single-Site Defective Term}

We first consider the effective Hamiltonian at $\gamma=g$ excluding
a diagonal constant, assuming periodic boundary conditions and $L>2$.
It is given by 
\begin{equation}
\hat{H}=\hat{H}_{\text{I}}-2g\hat{S}_{\mathcal{M}}^{-},\,\hat{S}_{\mathcal{M}}^{-}\equiv\frac{1}{2}\sum_{j\in\mathcal{M}}(\hat{\sigma}_{j}^{x}-i\hat{\sigma}_{j}^{y}).\label{eq:effective_Ising}
\end{equation}
If the measurement is applied to a single local site, $\mathcal{M}=\{m\}\subseteq\{1,2,\cdots,L\}$,
it is found that within the subspace of $\hat{H}_{\text{I}}$ spanned
by $\{|2;m-1\rangle,|2;m-1,m\rangle\}$, the $\hat{H}$ has a matrix
representation given by 
\begin{equation}
\left(\begin{array}{cc}
J(4-L) & -2g\\
0 & J(4-L)
\end{array}\right).
\end{equation}
Noticing that $\{|2;m-1\rangle,|2;m-1,m\rangle\}$ is decoupled to
other subspaces of $\hat{H}_{\text{I}}$, we find that the matrix
representation of $\hat{H}$ takes the block-diagonal form 
\begin{equation}
H=H^{\prime}\oplus\left(\begin{array}{cc}
J(4-L) & -2g\\
0 & J(4-L)
\end{array}\right),
\end{equation}
where $H^{\prime}$ represents the remaining $(2^{L}-2)\times(2^{L}-2)$
matrix sub-block.

The matrix $H$ can be further transformed to its Jordan normal form
using an invertible matrix $S$, 
\begin{equation}
S^{-1}HS=J_{H}^{\prime}\oplus J_{2}(J(4-L)),
\end{equation}
where $J_{H}^{\prime}$ is the $(2^{L}-2)\times(2^{L}-2)$ sub-block
of the remaining matrix. The Jordan block $J_{2}(J(4-L))$ signals
the presence of an EP of order $2$ in the effective Hamiltonian $\hat{H}$
or $\hat{H}_{\mathrm{TI}}^{\mathrm{eff}}(\gamma=g)$.

\subsubsection{Continuous Many-Site Defective Term}

\begin{figure}
\includegraphics[width=3.3in]{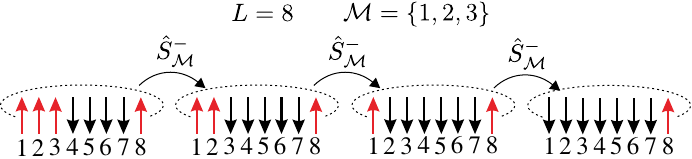}

\caption{\label{fig:Ising_EP_subspace}Symmetry for the translation of the
domain wall and the effect of the defective term $\hat{S}_{\mathcal{M}}^{-}$.
The system (\ref{eq:effective_Ising}) here is considered for $L=8$
and $\mathcal{M}=\{1,2,3\}$. Four periodic boundary spin chains from
the left to the right represent the states $|2;8,1,2,3\rangle$, $|2;8,1,2\rangle$,
$|2;8,1\rangle$, $|2;8\rangle$, respectively. These states have
the same number of domain wall $D=2$, and thus the same eigenenergy
$E_{D}=J(D-L)$ with respect to the Ising Hamiltonian $\hat{H}_{I}$.
The defective term $\hat{S}_{\mathcal{M}}^{-}$ acts as the domain
wall translation operator within this subspace.}
\end{figure}

We now consider the case where the measurement is applied to $|\mathcal{M}|$
continuous sites, $\mathcal{M}=\{m,m+1,\cdots,m+|\mathcal{M}|-1\}\subsetneq\{1,2,\cdots,L\}$,
with $L-2\geq|\mathcal{M}|\geq2$. Within the $|\mathcal{M}|+1$ dimensional
subspace of $\hat{H}_{\text{I}}$ spanned by $\{|2;m-1\rangle,|2;m-1,m\rangle,\cdots,|2;m-1,m,\dots,m+|\mathcal{M}|-1\rangle\}$,
the matrix representation of $\hat{H}$ is as follows 
\begin{equation}
\mathbb{J}_{|\mathcal{M}|+1}=\left(\begin{array}{ccccc}
J(4-L) & -2g & 0 & \cdots & 0\\
0 & J(4-L) & -2g & \cdots & 0\\
\vdots & \vdots & \ddots & \ddots & \vdots\\
0 & 0 & 0 & \ddots & -2g\\
0 & 0 & 0 & \cdots & J(4-L)
\end{array}\right).
\end{equation}
The defective term $\hat{S}_{\mathcal{M}}^{-}$ acts as a ``ladder''
operator in this degenerate subspace of $\hat{H}_{\text{I}}$. As
an example, consider $L=8$, and $\mathcal{M}=\{1,2,3\}$. The subspace
spanned by $\{|2;8,1,2,3\rangle,|2;8,1,2\rangle,|2;8,1\rangle,|2;8\rangle\}$
is generated by the translation of the domain wall from state $|2;8\rangle$.
As shown in Fig.~\ref{fig:Ising_EP_subspace}, the defective term
in the subspace transforms the states from left to right.

If the Hamiltonian contains only the matrix $\mathbb{J}_{|\mathcal{M}|+1}$,
we can readily bring it to the Jordan normal form $J_{|\mathcal{M}|+1}(J(4-L))$
and find the EP of order $|\mathcal{M}|+1$. However, things become
slightly more complicated than the case of $|\mathcal{M}|=1$, as
the aforementioned subspace will couple with other subspaces, implying
that the above block is coupled with other blocks, i.e.,
\begin{equation}
H=\left(\begin{array}{cc}
\left(H^{\prime}\right) & \left(\begin{array}{c}
\text{cross terms}\end{array}\right)\\
\left(\begin{array}{c}
\text{cross terms}\end{array}\right) & \left(\mathbb{J}_{|\mathcal{M}|+1}\right)
\end{array}\right),
\end{equation}
where $\left(H^{\prime}\right)$ denotes the remaining $(2^{L}-|\mathcal{M}|-1)\times(2^{L}-|\mathcal{M}|-1)$
matrix sub-block. While in principle we could follow the procedure
outlined in the above Heisenberg case (using elementary operations
and counting the dimension image space to find the rank and the Jordan
normal form), the Ising model does not provide us with a concise set
of indices to label the states, like $n$, $s_{n}$ and $m_{s}$ in
the Heisenberg case, rendering the process tedious. As a result, we
prefer to provide numerical evidence for our findings. Specifically,
the effective Hamiltonian $\hat{H}_{\mathrm{TI}}^{\mathrm{eff}}(\gamma=g)$
with a periodic boundary condition and $L>2$ possesses an EP of order
$|\mathcal{M}|+1$ if the measurement is applied to $|\mathcal{M}|$
continuous sites satisfying $L-2\geq|\mathcal{M}|\geq1$. This results
can be repeated by the following Mathematica codes \citep{WMathematicaOnline}.

\begin{widetext}%
\noindent\fbox{\begin{minipage}[t]{1\columnwidth - 2\fboxsep - 2\fboxrule}%
These codes define the Hamiltonian: 
\begin{quote}
H{[}J\_,g\_,L\_,M\_{]}:=Module{[}\\
\{KP=KroneckerProduct, PM=PauliMatrix, KPn=Fold{[}KroneckerProduct{]}{[}Table{[}\#,\#2{]}{]}\&,
KPM\}, \\
KPM{[}a\_,b\_,l\_,m\_{]}:=Sum{[}If{[}l>m{[}{[}i{]}{]}>1,KP{[}KPn{[}b,m{[}{[}i{]}{]}-1{]},a,KPn{[}b,l-m{[}{[}i{]}{]}{]}{]},If{[}m{[}{[}i{]}{]}==1,KP{[}a,KPn{[}b,l-1{]}{]},If{[}m{[}{[}i{]}{]}==l,
KP{[}KPn{[}b,l-1{]},a{]}{]}{]}{]},\{i,Length{[}m{]}\}{]};\\
-J{*}(KPM{[}KP{[}PM{[}3{]},PM{[}3{]}{]},PM{[}4{]},L-1,Range{[}L-1{]}{]}+KP{[}PM{[}3{]},KPn{[}PM{[}4{]},L-2{]},PM{[}3{]}{]})-g{*}KPM{[}PM{[}1{]}-I{*}PM{[}2{]},PM{[}4{]},L,M{]}{]};
\end{quote}
These codes count the order of EPs: 
\begin{quote}
EPsOrder:=Module{[}\{tmp=Reverse{[}SortBy{[}SplitBy{[}Total{[}JordanDecomposition{[}H{[}J,g,L,M{]}{]}{[}{[}2{]}{]}/.\{J->0\},\{2\}{]},1{]},Last{]}{]}\},Table{[}If{[}Total{[}tmp{[}{[}i{]}{]}{]}>0,Total{[}tmp{[}{[}i{]}{]}{]}+1,0{]},\{i,Length{[}tmp{]}\}{]}/.
\{x\_\_\_,0..\}:>\{x\}{]};
\end{quote}
These codes provide numerical evidences for our results: 
\begin{quote}
Do{[}L=i;M=Range{[}1,j{]};Print{[}\textquotedbl L=\textquotedbl ,i,\textquotedbl ,
|\textbackslash{[}ScriptCapitalM{]}|=\textquotedbl ,j,\textquotedbl ,
Order of EPs: \textquotedbl ,OEP=EPsOrder,\textquotedbl -{}-{}-{}-{}-{}-{}-{}-{}-{}-{}-{}-{}-{}-{}-{}-{}-{}-{}-{}-{}-{}-Satisfy
EP(|\textbackslash{[}ScriptCapitalM{]}|+1) for |\textbackslash{[}ScriptCapitalM{]}|<=L-2,
EP(L-1) for |\textbackslash{[}ScriptCapitalM{]}|>L-2? \textquotedbl ,If{[}j<=i-2,OEP{[}{[}1{]}{]}==j+1,OEP{[}{[}1{]}{]}==i-1{]}{]},\{i,Range{[}3,7{]}\},\{j,Range{[}1,i{]}\}{]};
\end{quote}
\end{minipage}}\end{widetext}

\section{Generalized Eigenvectors and Dynamical Scaling}

At the exceptional point (EP), the non-Hermitian Hamiltonian becomes
non-diagonalizable, leading to the coalescence of two or more energy
eigenstates. This phenomenon can be observed by tracking the asymptotic
behaviors of distinct eigenstates near the EP. The coalescence of
eigenvectors implies a reduction in the dimension of the Hilbert space
of the non-Hermitian Hamiltonian operator at the EP. However, despite
this reduction, state vectors may still retain information about these
lost dimensions. Consequently, the dynamical evolution of states generated
by the non-diagonalizable Hamiltonian appears unusual, as the eigenstates
are insufficient to expand this state. To address this issue, we can
construct generalized eigenvectors. The introduction of these vectors
allows us to build a set of linearly independent basis vectors to
expand any states. In the following sections, we will review the concept
of generalized eigenvectors as presented in \citep{bronson1991matrix},
derive the evolution equation given in the main text (Eq.~(4)), and
use it to derive the integer family of dynamical scaling in Ising
and Heisenberg systems under monitoring.

\subsection{Generalized Eigenvectors}

The concept of generalized eigenvectors and their primary properties
are encapsulated in the following two definitions and three theorems
\citep{bronson1991matrix}. 

\emph{Definition 1}. A vector $|V_{n}\rangle$ is a generalized eigenvector
of rank $n$ corresponding to Hamiltonian $\hat{H}$ and eigenvalue
$\lambda$ if $(\hat{H}-\lambda\hat{I})^{n}|V_{n}\rangle=0$ but $(\hat{H}-\lambda\hat{I})^{n-1}|V_{n}\rangle\neq0$,
where $\hat{I}$ is the identity operator.

Note that the rank-1 generalized eigenvector is the usual eigenvector.

\emph{Definition 2}. The chain generated by $|V_{n}\rangle$ is the
set of vectors $\{|V_{n}\rangle,|V_{n-1}\rangle,\cdots,|V_{1}\rangle\}$
given by the sequence 
\begin{align}
|V_{n}\rangle & =(\hat{H}-\lambda\hat{I})^{-1}|V_{n-1}\rangle=(\hat{H}-\lambda\hat{I})^{-2}|V_{n-2}\rangle\nonumber \\
 & =...=(\hat{H}-\lambda\hat{I})^{1-n}|V_{1}\rangle,
\end{align}
where $|V_{m}\rangle,(m=1,2,\cdots,n-1)$ are by definition $m$-rank
generalized eigenvectors corresponding to Hamiltonian $\hat{H}$ and
eigenvalue $\lambda$.

\emph{Theorem 1}. $|V_{m}\rangle$ is a generalized eigenvector of
rank $m$ corresponding to Hamiltonian $\hat{H}$ and eigenvalue $\lambda$.

\emph{Theorem 2}. A chain is a linearly independent set of vectors.

\emph{Theorem 3}. Every $n\times n$ matrix possesses $n$ linearly
independent generalized eigenvectors.

Theorem 3 implies that we can always use the generalized eigenvector
of a Hamiltonian at the EP to study dynamical evolutions by expanding
$|\psi\rangle=\sum_{m}a_{m}|E_{m}\rangle+\sum_{j=1}^{N_{\mathrm{EP}}}\sum_{n=1}^{\mathscr{O}_{j}}c_{n}^{(j)}|V_{n}^{(j)}\rangle$,
where each term is explained in the main text. Now, let's prove the
evolution equation for the generalized eigenvector of rank $n$ given
in Eq.~(4). Since $\hat{H}$ and $\lambda\hat{I}$ commute, we can
write the binomial formula $(\hat{H}+\lambda\hat{I})^{k}=\sum_{m=0}^{k}C_{k}^{m}\lambda^{k-m}\hat{H}^{m}$,
yielding 
\begin{equation}
\hat{H}^{k}=\sum_{m=0}^{k}C_{k}^{m}\lambda^{k-m}(\hat{H}-\lambda\hat{I})^{m},
\end{equation}
where $C_{k}^{m}=\frac{k!}{m!(k-m)!}$. Multiplying by $|V_{n}\rangle$,
we obtain 
\begin{align}
\hat{H}^{k}|V_{n}\rangle & =\sum_{m=0}^{n-1}\frac{1}{m!}\left(\frac{k!}{(k-m)!}\lambda^{k-m}\right)\left((\hat{H}-\lambda\hat{I})^{m}|V_{n}\rangle\right)\nonumber \\
 & =\sum_{m=0}^{n-1}\frac{1}{m!}\frac{d^{m}\lambda^{k}}{d\lambda^{m}}|V_{n-m}\rangle.
\end{align}
Considering the polynomial function of $\hat{H}$, $e^{-i\hat{H}t}=\sum_{k=0}^{\infty}\frac{(-it)^{k}}{k!}\hat{H}^{k}$,
we find 
\begin{align}
e^{-iHt}|V_{n}\rangle & =\sum_{m=0}^{n-1}\frac{1}{m!}\frac{d^{m}}{d\lambda^{m}}\left(\sum_{k=0}^{\infty}\frac{(-it)^{k}}{k!}\lambda^{k}\right)|V_{n-m}\rangle\nonumber \\
 & =\sum_{m=0}^{n-1}\frac{1}{m!}\frac{d^{m}}{d\lambda^{m}}e^{-i\lambda t}|V_{n-m}\rangle\nonumber \\
 & =e^{-i\lambda t}\sum_{m=0}^{n-1}\frac{(-it)^{m}}{m!}|V_{n-m}\rangle,
\end{align}
which is Eq.~(4). It is evident that for an EP of order $n$, there
exists a generalized eigenvector of rank $n$, which carries a $t^{n-1}$
in the dynamics. Therefore, as explained in the main text, the expectation
value of generic observables $\langle\psi(t)|\hat{O}|\psi(t)\rangle$
should exhibit the dynamical scaling $\propto t^{2(O_{j_{\mathrm{max}}}-1)}$
in the late-time dynamics provided $\langle V_{1}^{(j_{\mathrm{max}})}|\hat{O}|V_{1}^{(j_{\mathrm{max}})}\rangle\neq0$. 

\subsection{Dynamical Scaling in Ising and Heisenberg Models}

In Sec.\ref{sec:Higher-order-EPs}, we demonstrated the existence
of the EP of order $L+1$ for an arbitrary $|\mathcal{M}|$ in the
effective Heisenberg model with Hamiltonian $\hat{H}_{\mathrm{TH}}^{\text{eff}}$.
Notably, the rank-1 generalized eigenvector corresponding to eigenvalue
$E_{0}$ is $|\downarrow\downarrow\cdots\downarrow\rangle$, i.e.,
$|0,s_{0},-s_{0}\rangle$ in Eq.(\ref{eq:Heisenberg_EP_eigenvector}).
Taking the expectation of the total magnetization along the $z$-direction,
we find 
\begin{equation}
\langle\downarrow\downarrow\cdots\downarrow|\left(\sum_{j=1}^{L}\frac{\hat{\sigma}_{j}^{z}}{2}\right)|\downarrow\downarrow\cdots\downarrow\rangle\neq0,
\end{equation}
which explains the dynamical power law scaling $t^{2L}=t^{2(L+1-1)}$
in the late-time. As for the total magnetization along the other two
directions, we find 
\begin{align}
\langle\downarrow\downarrow\cdots\downarrow|\left(\sum_{j=1}^{L}\frac{\hat{\sigma}_{j}^{x}}{2}\right)|\downarrow\downarrow\cdots\downarrow\rangle & =0,\\
\langle\downarrow\downarrow\cdots\downarrow|\left(\sum_{j=1}^{L}\frac{\hat{\sigma}_{j}^{y}}{2}\right)|\downarrow\downarrow\cdots\downarrow\rangle & =0,
\end{align}
which imply that the maximal scaling exponents in their late-time
dynamics depend on the generalized eigenvectors of higher ranks. Nevertheless,
their scaling behaviors take the form $t^{2L-q}$ with some positive
integer $q$, if the maximal rank $n_{\max}$ of the generalized eigenvectors
in the relevant chain grow with system size at the same rate, $n_{\max}\propto L$.

In Sec.~\ref{sec:Higher-order-EPs}, we also demonstrated the existence
of the EP of order $|\mathcal{M}|+1$ in the effective Hamiltonian
$\hat{H}_{\mathrm{TI}}^{\text{eff}}$ with periodic boundary condition
$\mathcal{M}=\{m,m+1,\cdots,m+|\mathcal{M}|-1\}$ and $L-2\geq|\mathcal{M}|\geq1$.
Notably, one of the rank-1 generalized eigenvectors corresponding
to eigenvalue $J(4-L)$ is $|\downarrow_{1}\cdots\downarrow_{m-2}\uparrow_{m-1}\downarrow_{m}\cdots\downarrow_{L}\rangle$,
in which $\sum_{j=1}^{L}\frac{\hat{\sigma}_{j}^{z}}{2}$ has a non-vanishing
expectation value. This explains the dynamical power law scaling $t^{2|\mathcal{M}|}$
in the late-time. Similarly, the expectation values for the total
magnetization along the other two directions in $|\downarrow_{1}\cdots\downarrow_{m-2}\uparrow_{m-1}\downarrow_{m}\cdots\downarrow_{L}\rangle$
are zeros, and thus their scaling behaviors take the form $t^{2|\mathcal{M}|-q}$
with some positive integer $q$, if the maximal rank $n_{\max}$ of
the generalized eigenvectors in the relevant chain grow with system
size at the same rate, $n_{\max}\propto|\mathcal{M}|$.

\section{Perturbation Theory and Diverging Timescale}

The quantum dynamics discussed in our main text can be encapsulated
effectively by a non-Hermitian Hamiltonian. It is reasonable to conjecture
that the diverging characteristic timescales, as depicted in Figs.~2(b1-b2),
near $\gamma=g$ could be a manifestation of the energy splitting
in the proximity of the exceptional point (EP). In this section, we
first provide a concise review of the perturbation theory near the
EP \citep{kato2013perturbation,knopp2013theory}, and then demonstrate
that the diverging characteristic timescale $\tau\propto|\gamma-g|^{-1/2}$,
found in the main text, can indeed be explained by the energy splitting
near the EP.

\subsection{Perturbation Theory}

For simplicity, let's consider a Hamiltonian in a $d$-dimensional
Hilbert space 

\begin{equation}
\hat{H}(\epsilon)\equiv\hat{H}_{1}+\epsilon\hat{H}_{2},\,\epsilon\in\mathbb{C}.
\end{equation}
We assume that this Hamiltonian has only one EP of order $d$ at $\epsilon=0$,
where all the eigenvectors coalesce and the only eigenvalue is $E(0)$.
We also assume that the $d$ eigenvalues at $\epsilon\neq0$ are distinct,
i.e., $E_{k}(\epsilon)\neq E_{n}(\epsilon)$ for $k,n\in\{1,2,\dots,d\}$
and $k\neq n$. 

Now, let's restrict $\epsilon$ to be a nonzero value in the complex
plane and denote the set of eigenvalues as

\begin{equation}
\{E_{1}(\epsilon),E_{2}(\epsilon),\dots,E_{d}(\epsilon)\},\label{eq:eigenenergies_near_EP}
\end{equation}
These eigenvalues are uniquely specified through the characteristic
function

\begin{equation}
\det(\hat{H}(\epsilon)-E_{n})=0.
\end{equation}
Next, we move the perturbation $\epsilon$ starting from $z_{0}$,
through a circle with the center being the origin, back to $z_{0}$
again. The $d$ eigenvalues in this process can be analytically continued.
When back to the same point $z_{0}$, the $d$ eigenvalues must either
be unchanged or undergo a cyclic permutation. Without loss of generality,
we assume

\begin{equation}
\{E_{1}(\epsilon),\dots,E_{p}(\epsilon)\},\,2\leq p\leq d\label{eq:EP_cycle}
\end{equation}
is a cycle of period $p$, which means that when $\epsilon$ is moved
one circle, the permutation carries $E_{1}$ into $E_{2}$, $\dots$,
$E_{p-1}$ into $E_{p}$, and $E_{p}$ into $E_{1}$. For a given
$h\in\{1,\dots p\}$, we find $E_{h}(\epsilon^{p})$ is a regular,
single-valued function of $\epsilon$, which allows us to develop
the Laurent series, $E_{h}(\epsilon^{p})=\sum_{m=-\infty}^{\infty}c_{m}\epsilon^{m}$,
and thus

\begin{equation}
E_{h}(\epsilon)=\sum_{m=-\infty}^{\infty}c_{m}\epsilon^{m/p}.
\end{equation}
The fact that all eigenvalues converge to $E(0)$ as $\epsilon\rightarrow0$
implies that there is no negative power of $\epsilon$, i.e., $E_{h}(\epsilon)=E(0)+\sum_{m=1}^{\infty}c_{m}\epsilon^{m/p}$.
Finally, the permutation indicates the following form

\begin{equation}
E_{h}(\epsilon)=E(0)+\sum_{m=1}^{\infty}c_{m}e^{2\pi im(h-1)/p}\epsilon^{m/p},\label{eq:series_energies}
\end{equation}
where $h\in\{1,\dots p\}$.

Interestingly, there will be additional constraints for a system where
$\epsilon>0$ and $\epsilon<0$ exhibit parity-time reversal symmetry
preserving and broken phase, respectively. Specifically, $\hat{H}(\epsilon>0)$
and $\hat{H}(\epsilon<0)$ have complete real spectra and complex
conjugated spectra, respectively. First, the period of cycle in (\ref{eq:EP_cycle})
must exhibit $p>1$, since if $p=1$, the energies in Eq.~(\ref{eq:series_energies})
cannot be real and complex conjugate for $\epsilon>0$ and $\epsilon<0$,
respectively. Second, the period $p$ can only be $2$ as the eigenvalues
for $\epsilon>0$ are all real. Therefore, for every pair of eigenvalues
$E_{\pm}$ in the period of $2$, they simply satisfy

\begin{equation}
E_{\pm}(\epsilon)=E(0)\pm\sum_{m=1}^{\infty}c_{m}\epsilon^{m/2}.
\end{equation}
Especially, for a small perturbation $\epsilon$, the gap $\Delta_{E}\equiv E_{+}-E_{-}$
is found to be

\begin{equation}
\Delta_{E}\propto\epsilon^{1/2}
\end{equation}
with an exponent $1/2$.

\subsection{Parity-Time Reversal Symmetry Transition and the Diverging Timescale}

\begin{figure*}
\includegraphics[width=2.2in]{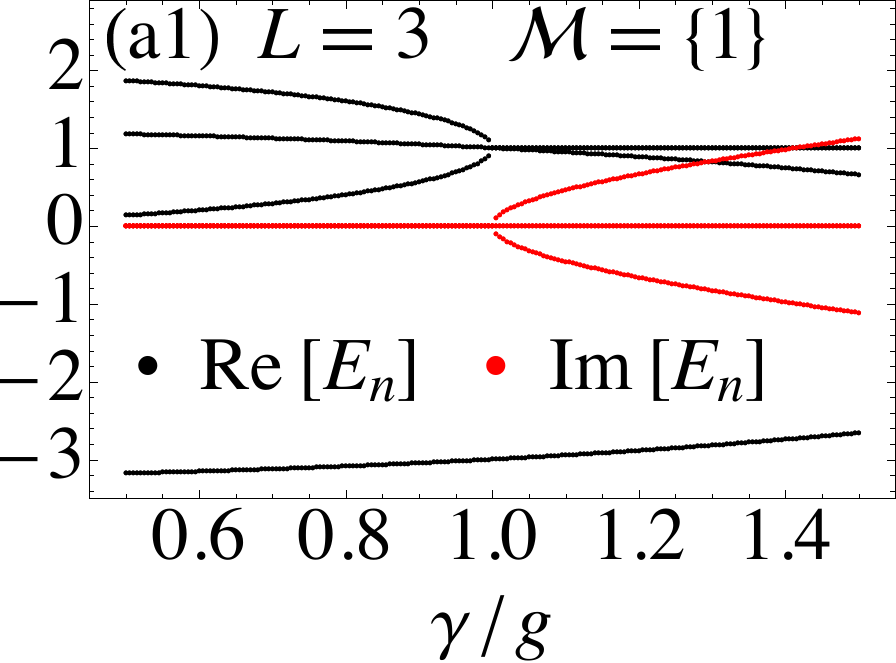}~\includegraphics[width=2.2in]{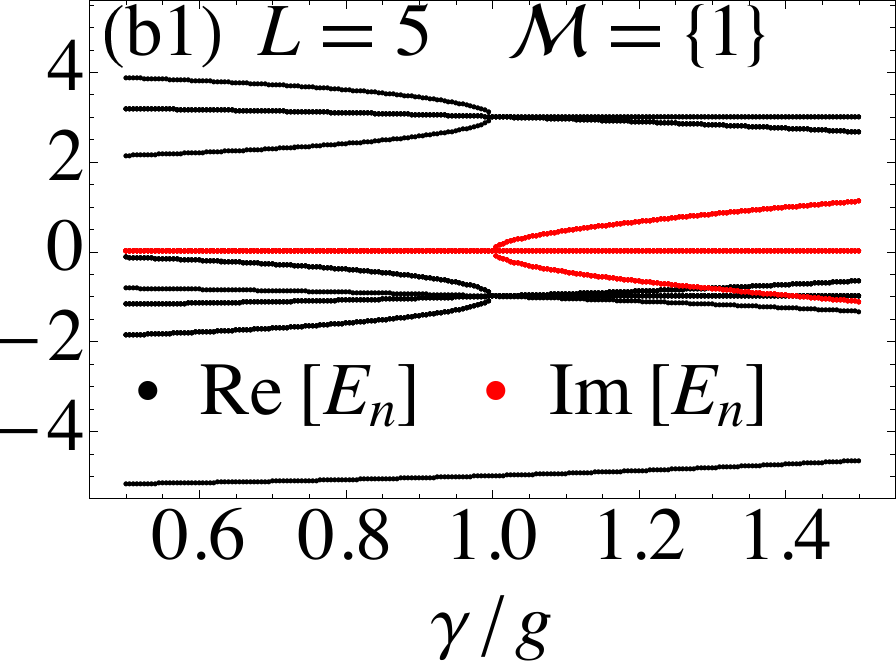}~\includegraphics[width=2.2in]{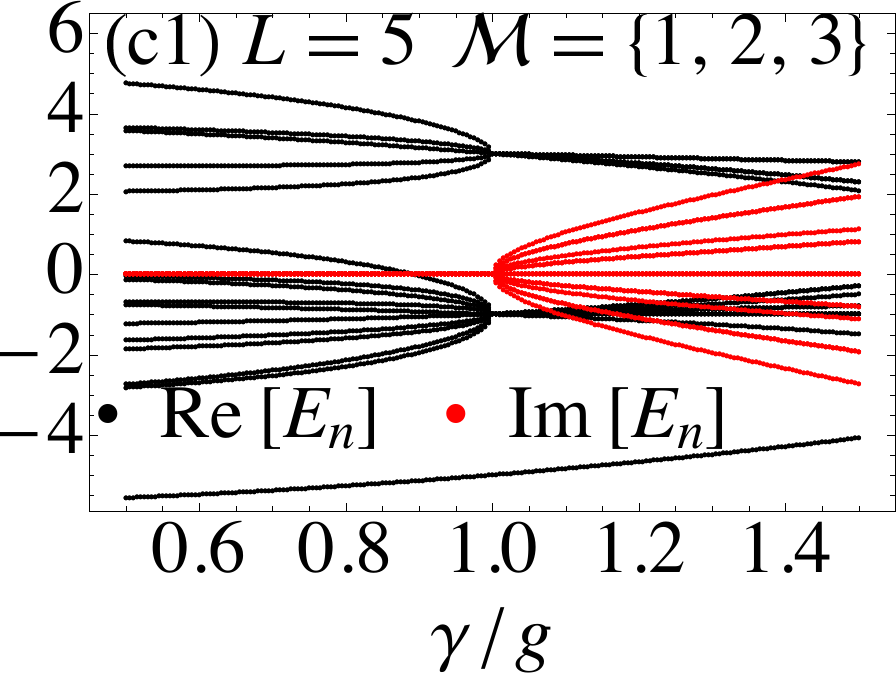}

\includegraphics[width=2.2in]{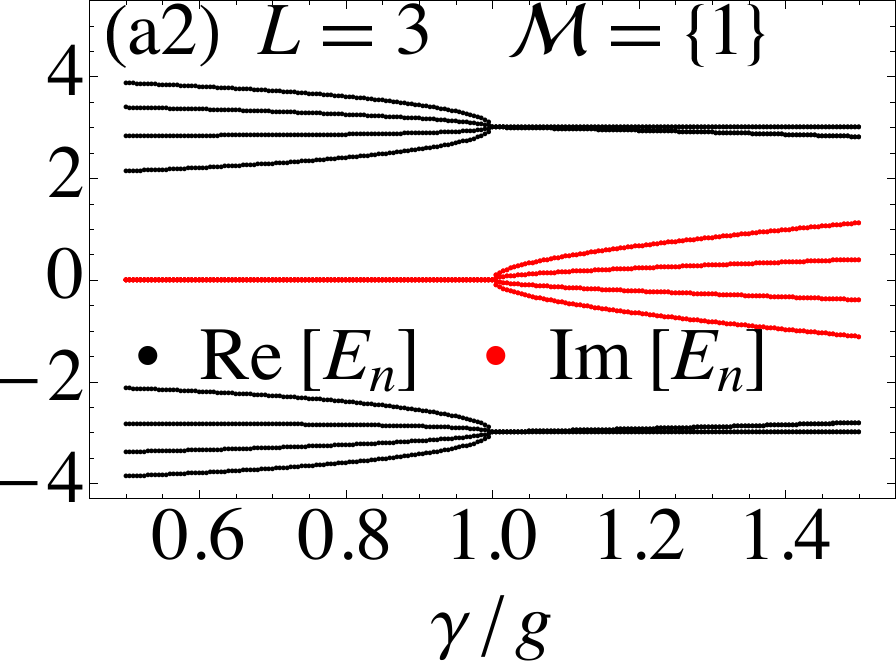}~\includegraphics[width=2.2in]{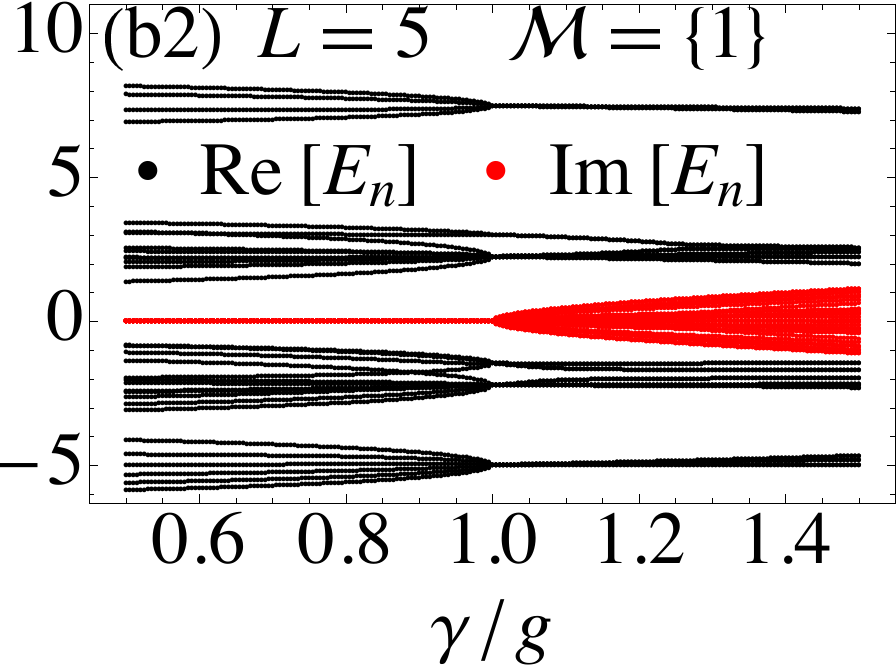}~\includegraphics[width=2.2in]{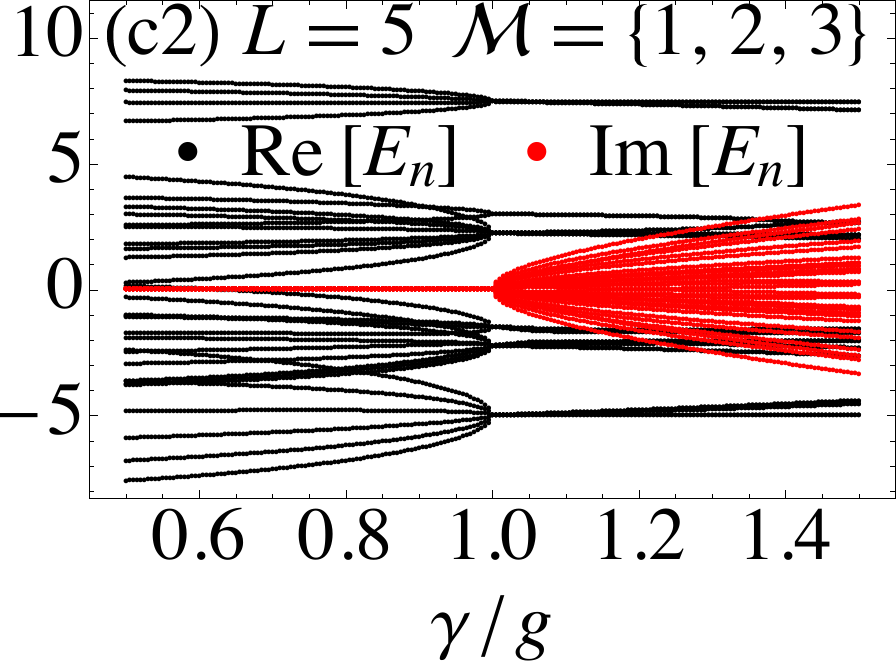}\caption{\label{fig:Spectra}Spectra of effective Ising (Eq.~(\ref{eq:Eff_Ising_Prime}))
and Heisenberg models (Eq.~(\ref{eq:Eff_Heisenberg_Prime})). The
spectra of the effective Ising model (a1-c1) and the effective Heisenberg
model (a2-c2) are calculated for different system sizes ($L=3,5$)
and measurement regions ($|\mathcal{M}|=1,3$ with $\mathcal{M}=\{1,2,\cdots,|\mathcal{M}|\}$).
All these eigenenergies $\{E_{n}\}$ are completely real for $\gamma<g$
and contain complex conjugated imaginary components for $\gamma>g$.}
\end{figure*}

The effective Ising and Heisenberg models have the Hamiltonians 
\begin{align}
\hat{H}_{\mathrm{TI}} & =-J\sum_{j=1}^{L}\hat{\sigma}_{j}^{z}\hat{\sigma}_{j+1}^{z}-\sum_{j\in\mathcal{M}}(g\hat{\sigma}_{j}^{x}-i\gamma\hat{\sigma}_{j}^{y}),\label{eq:Eff_Ising_Prime}\\
\hat{H}_{\mathrm{TH}} & =-J\sum_{j=1}^{L}\sum_{a=x,y,z}\hat{\sigma}_{j}^{a}\hat{\mathbf{\sigma}}_{j+1}^{a}-\sum_{j\in\mathcal{M}}(g\hat{\sigma}_{j}^{x}-i\gamma\hat{\sigma}_{j}^{y}),\label{eq:Eff_Heisenberg_Prime}
\end{align}
For simplicity, we have dropped a trivial constant in both effective
Hamiltonians, i.e., $\hat{H}_{\mathrm{TI}}=\hat{H}_{\mathrm{TI}}^{\text{eff}}+i\gamma|\mathcal{M}|$
and $\hat{H}_{\mathrm{TH}}=\hat{H}_{\mathrm{TH}}^{\text{eff}}+i\gamma|\mathcal{M}|$.
We find that the energy spectra in both systems at $\gamma<g$ ($\gamma>g$)
are completely real (all complex conjugate pairs) regardless of the
system sizes $L$ and the measurement regions $|\mathcal{M}|$, as
shown in Fig.~\ref{fig:Spectra}. This indicates that the EP ($\gamma=g$)
serves as the parity-time reversal symmetry transition point. 

The parity-time reversal symmetry transition becomes evident if we
explicit write the parity $\mathcal{P}$ and time $\mathcal{T}$ operators
as $\mathcal{P}=\otimes_{j=1}^{L}\hat{\sigma}_{j}^{y}$, $\mathcal{T}=(\otimes_{j=1}^{L}\hat{\sigma}_{j}^{y})\mathcal{K}$,
where $\mathcal{K}$ generates the Hermitian conjugation. The Hamiltonian
$\hat{H}_{\mathrm{TI}}$ and $\hat{H}_{\mathrm{TH}}$ are $\mathcal{PT}$
symmetric, i.e., $\mathcal{PT}\hat{H}\left(\mathcal{PT}\right)^{-1}=\hat{H}$
for $\hat{H}=\hat{H}_{\mathrm{TI}},\hat{H}_{\mathrm{TH}}$. The eigenstates
for these Hamiltonian at $\gamma<g$ are also invariant under the
action of $\mathcal{PT}$, while at $\gamma>g$ changes to their Hermitian
conjugation. This means that the parity-time reversal symmetries of
these systems are spontaneously broken at $\gamma>g$. 

Both the effective Ising and Heisenberg Hamiltonians have $2^{L}$
eigenenergies in the vicinities of $\gamma=g$, denoted as $\{E_{n,1}\}\cup\{E_{n,\pm}\}$,
where $E_{n,1}$ does not involve a permutation (period $p=1$ in
(\ref{eq:EP_cycle})) and $E_{n,\pm}$ involve a permutation ($p=2$)
as discussed above. Consider a general state expanded as

\begin{equation}
|\psi\rangle=\sum_{n}c_{n,1}|E_{n,1}\rangle+\sum_{n}(c_{n,+}|E_{n,+}\rangle+c_{n,-}|E_{n,-}\rangle).
\end{equation}
The dynamics of the expectation value for an arbitrary operator that
is non-diagonal in the energy representation is given by $\langle\psi|e^{it\hat{H}}\hat{O}e^{-it\hat{H}}|\psi\rangle$,
where the slowest modes are given by coefficients like

\begin{equation}
e^{\pm it(E_{n,+}-E_{n,-})}.
\end{equation}
Especially, for $-1\ll\frac{\gamma-g}{g}<0$, the characteristic
timescale $\tau$ in the slowest mode is fitted by the oscillation
function $\cos(t/\tau)$, and exhibits 
\begin{equation}
\tau\propto1/|E_{n,+}-E_{n,-}|\propto|\gamma-g|^{-1/2}.
\end{equation}
For $0<\frac{\gamma-g}{g}\ll1$, the characteristic timescale $\tau$
in the slowest mode is fitted by the exponential function $e^{t/\tau}$,
and also exhibits 
\begin{equation}
\tau\propto1/|E_{n,+}-E_{n,-}|\propto|\gamma-g|^{-1/2}.
\end{equation}

\bibliographystyle{apsrev4-1}

\end{document}